\documentclass[dvipsnames]{aa} 

\usepackage{graphicx}
\usepackage{txfonts}
\usepackage{stfloats}
\usepackage[colorlinks=true,linkcolor=blue,citecolor=blue,urlcolor=blue]{hyperref}
\begin{document}

    \title{HYACINTH: HYdrogen And Carbon chemistry in the INTerstellar medium in Hydro simulations} 

   \author{
        Prachi Khatri \inst{1}
        \and Cristiano Porciani\inst{1,2,3,4} 
        \and Emilio Romano-Díaz\inst{1} 
        \and Daniel Seifried\inst{5} 
        \and Alexander Sch\"abe\inst{6}
        }

   \institute{Argelander Institute f\"ur Astronomie, Auf dem H\"ugel 71, D-53121 Bonn, Germany,\\
             \email{pkhatri@astro.uni-bonn.de }
             \and
             SISSA, International School for Advanced Studies, Via Bonomea
            265, 34136 Trieste TS, Italy
            \and
            Dipartimento di Fisica - Sezione di Astronomia, Universit\`a di Trieste, Via Tiepolo 11, 34131 Trieste, Italy
            \and
            IFPU, Institute for Fundamental Physics of the Universe, via Beirut 2, 34151 Trieste, Italy
             \and
             Universit\"at zu K\"oln, I. Physikalisches Institut, Z\"ulpicher Str 77, D-50937 K\"oln, Germany
             \and
             T\"UV NORD EnSys GmbH \& Co. KG, Am T\"UV 1, D-30519 Hannover
             }

   \date{Received 16 February 2024 / Accepted 13 June 2024}

    \titlerunning{HYACINTH}
    \authorrunning{P. Khatri et al.}
% \abstract{}{}{}{}{} 
% 5 {} token are mandatory
 
  \abstract
   {}
   {We present a new sub-grid model, HYACINTH -- HYdrogen And Carbon chemistry in the INTerstellar medium in Hydro simulations -- for computing the non-equilibrium abundances of ${\rm H_2}$ and its carbon-based tracers, namely ${\rm CO}$, ${\rm C}$, and ${\rm C^+}$, in cosmological simulations of galaxy formation.}
   {The model accounts for the unresolved density structure in simulations using a variable probability distribution function of sub-grid densities and a temperature-density relation. Included is a simplified chemical network that has been tailored for hydrogen and carbon chemistry within molecular clouds and easily integrated into large-scale simulations with minimal computational overhead. As an example, we applied HYACINTH to a simulated galaxy at redshift $z \sim 2.5$ in post-processing and compared the resulting abundances with observations.}
   {The chemical predictions from HYACINTH are in reasonable agreement with high-resolution molecular-cloud simulations at different metallicities. By post-processing a galaxy simulation with HYACINTH, we reproduced the ${\rm H}\,\rm \textsc{i}-{\rm H_2}$ transition as a function of the hydrogen column density $N_{\rm H}$ for both Milky-Way-like and Large-Magellanic-Cloud-like conditions. We also matched the $N_{{\rm CO}}$ versus $N_{{\rm H_2}}$ relation inferred from absorption measurements towards Milky-Way molecular clouds, although most of our post-processed regions occupy the same region as (optically) dark molecular clouds in the $N_{{\rm CO}} - N_{\rm H_2}$ plane. Column density maps reveal that ${\rm CO}$ is concentrated in the peaks of the ${\rm H_2}$ distribution, while atomic carbon more broadly traces the bulk of ${\rm H_2}$ in our post-processed galaxy. Based on both the column density maps and the surface density profiles of the different gas species in the post-processed galaxy, we find that ${\rm C^+}$ maintains a substantially high surface density out to $\sim 10 \, \rm kpc$ as opposed to other components that exhibit a higher central concentration. This is similar to the extended $[\rm C \, \rm \textsc{ii}]$ emission found in some recent observations at high redshifts.
    }
   {}
 

   \maketitle
%
%-------------------------------------------------------------------
\section{Introduction}
\label{sec:intro}
Molecular gas plays a major role in the interstellar medium (ISM) of galaxies, providing the necessary conditions and likely serving as the primary fuel for star formation. The cosmic molecular gas density in the Universe, as inferred from blind surveys such as the VLA CO Luminosity Density at High Redshift\footnote{\href{http://coldz.astro.cornell.edu/}{http://coldz.astro.cornell.edu/}} \citep[COLDz,][]{riechers19} and the ALMA Spectroscopic Survey in the HUDF\footnote{\href{https://aspecs.info/}{https://aspecs.info/}} \citep[ASPECS, ][]{decarli19, aspecs}, increases by roughly an order of magnitude between redshifts $z \sim 6$ and $z \sim 2$. This is accompanied by a similar trend in the cosmic star formation rate density that reaches its peak value at $z \sim 3-2$, a period known as `cosmic noon' \citep[see][for a review]{madau-dickinson14, schreiber-wuyts20}.

Investigating the build-up of the molecular gas reservoir in galaxies and its cosmic evolution is therefore pivotal to our understanding of the star formation history and galaxy assembly in the Universe.  Molecular hydrogen (${\rm H_2}$) is the dominant molecular gas component in galaxies. However, because of the lack of a permanent dipole moment and high excitation temperatures ($T\gtrsim 500 \, \rm K$) for its ro-vibrational transitions, ${\rm H_2}$ does not emit light under typical conditions of molecular clouds ($T\lesssim 100 \, \rm K$). Therefore, the presence and mass of $\rm H_2$ are routinely inferred via emission from tracers such as dust, ${\rm CO}$, and atomic fine-structure lines.

The low-$J$ rotational transitions of ${\rm CO}$ are the most commonly used tracers of molecular gas in galaxies \citep[e.g.][]{dickman86, solomon91, downes98, solomon-bout05, tacconi06, tacconi08, tacconi10, daddi10a, daddi10b, genzel10, bolatto13}. The ${\rm CO}$-to-${\rm H_2}$ conversion factor $\alpha_{{\rm CO}}$ captures the relation between the observed ${\rm CO} \; J=1 \rightarrow 0$ luminosity of a galaxy and the underlying molecular gas mass. Alternatively, for spatially resolved observations within the Milky Way (MW) and nearby galaxies, $X_{{\rm CO}}$ relates the ${\rm CO}$ intensity ($W_{{\rm CO}}$) to the ${\rm H_2}$ column density ($N_{\rm H_2}$) along the line of sight. The variation of $X_{{\rm CO}}$ with physical conditions, such as metallicity, stellar surface density, galactocentric distance, among others, has been extensively investigated in the MW and nearby galaxies \citep[see][for a review]{bolatto13}. At higher redshifts, however, the limited number of galaxies observed in ${\rm CO}$ poses a challenge to studying the dependence of $\alpha_{{\rm CO}}$ on other galaxy properties. This is further complicated by the fact that the ${\rm CO} \; J=1 \rightarrow 0$ transition is not accessible at high redshifts and observers have to rely on higher-$J$ transitions to obtain an estimate for it. This requires knowledge about the ${\rm CO}$ excitation ladder, thereby introducing another systematic uncertainty in employing ${\rm CO}$ as a molecular gas tracer. Moreover, in some galaxies (e.g. low-metallicity dwarf galaxies), a large amount of ${\rm H_2}$ is not traced by ${\rm CO}$ emission and is referred to as ${\rm CO}$-dark molecular gas \citep{wolfire10, madden20}.

Other tracers of molecular gas such as the $[{\rm C}\, \textsc{ii}]$ fine-structure line suffer from similar systematic effects. The nature and origin of this line are highly debated as it can arise from multiple ISM phases and not all of the $[{\rm C}\, \textsc{ii}]$ emission of a  galaxy is associated with the molecular gas phase. The presence of a $[{\rm C}\, \textsc{ii}]$ halo in some galaxies extending $2-3$ times farther than their rest-frame UV emission  \citep[see e.g.][]{fujimoto20} hints at an extended ${\rm C^+}$ reservoir devoid of molecular gas. Additionally, the decrease in the $[{\rm C}\, \textsc{ii}]$/far-infrared (FIR) luminosity ratio with increasing FIR luminosity, known as the `$[{\rm C}\, \textsc{ii}]$ deficit', further complicates the use of this line as a reliable tracer of molecular gas across galaxies.

Atomic carbon has been proposed as another reliable tracer of molecular gas. The fine structure lines of atomic carbon are expected to trace the bulk of the molecular gas in galaxies \citep{papadopoulos04, matteo_ci, glover-clark16} and have been used both in the local Universe as well as at high redshifts  \citep[e.g.][]{gerin00, ikeda02, weiss03, weiss05, walter11, valentino18, henriquez-brocal22}. However, the use of these lines requires assumptions about the relative abundance of atomic carbon and molecular hydrogen. 

Cosmological simulations are a useful tool for investigating the reliability of molecular gas tracers under different ISM conditions and galaxy environments. However, simulating the molecular gas content of galaxies is challenging as it requires modelling the various physical and chemical processes happening on a wide range of (spatial and temporal) scales. On the one hand, it is necessary to simulate galaxies in realistic environments as their ISM is affected by outflows and gas accretion from the cosmic web. On the other hand, molecular-cloud chemistry is regulated by conditions on sub-parsec scales. Early attempts at modelling ${\rm H_2}$ assumed a local chemical equilibrium between ${\rm H_2}$ formation and destruction \citep{kmt08, kmt09, kmt10} and these models have found extensive application in cosmological simulations \citep{kuhlen12, kuhlen13, hopkins14, thompson14, lagos15, dave16}. Despite their wide use, these equilibrium models do not account for the dynamic nature of the ISM and the long formation timescale for ${\rm H_2}$ \citep{tielens_hollenback85}. For example, \cite{pelupessy09}, \cite{matteo_h2}, \cite{richings16}, \cite{pallottini17}, \cite{alex_h2}, and \cite{hu21} provide a detailed discussion on the effect of non-equilibrium ${\rm H_2}$ chemistry on the integrated properties of simulated galaxies. However, cosmological hydrodynamical simulations with on-the-fly computations of the non-equilibrium chemical abundances are rare \citep[e.g.][]{dobbs08, christensen12, matteo_h2, semenov18,lupi18, lupi19, alex_h2, katz22, hu23} and often restricted to individual galaxies. Some of these studies only include a non-equilibrium chemical network for ${\rm H}\,\rm \textsc{i}$ and ${\rm H_2}$ but not the tracers \citep{lupi20a, lupi20b}. 

Despite tremendous improvement in the spatial resolution of cosmological simulations over the last decade, the state-of-the-art today is still far from resolving the clumpy ISM in molecular clouds. A clumpy ISM would allow for pockets of high-density gas where ${\rm H_2}$ formation would be enhanced. This enhancement is missed by simulations as the density is uniform below their resolution scale.  \cite{gnedin09} accounted for these unresolved densities by enhancing the ${\rm H_2}$ formation rate by an effective clumping factor $C$ assuming a log-normal density distribution for the gas. This technique was later tested by \cite{micic12} in their numerical study on the effect of the nature of turbulence on ${\rm H_2}$ formation. They found that using a clumping factor systematically overpredicts the ${\rm H_2}$ formation rate in regions with a high molecular fraction ($f_{\rm H_2} \gtrsim 0.5$). \cite{christensen12} adapted this method in their smooth particle hydrodynamics (SPH) simulations to model the non-equilibrium abundance of ${\rm H_2}$ in a cosmological simulation of a dwarf galaxy. However, similar to \cite{micic12}, they cautioned that this approach works well for dwarf galaxies where fully molecular gas is rare but would need further modifications at high molecular fractions. Moreover, the finite resolution of simulations implies that the temperature tracked in simulations is an average over the resolution element, similar to  any other quantity followed explicitly. This average fails to capture the true heterogeneous temperature distribution that is essential for determining the rates of chemical reactions taking place in the ISM. 

To overcome these limitations, \citet[][hereafter T15]{matteo_h2} developed a sub-grid model that accounts for the unresolved density structure in simulations by assuming a log-normal probability distribution function (PDF) of sub-grid densities. They assigned a temperature to each sub-grid density in the PDF using a temperature-density relation from high-resolution simulations of molecular clouds \citep{glover07b}. They obtained the ${\rm H_2}$ abundance in each resolution element by evolving their chemical network at the sub-grid level and integrating over the density PDF. They found good agreement between different gas and stellar properties of their simulated galaxy at $z=2$ and observations of high-redshift galaxies. 

In this paper, we improve upon the work of T15 and introduce a new sub-grid model, HYACINTH -- HYdrogen And Carbon chemistry in the INTerstellar medium in Hydro simulations -- for on-the-fly computation of the non-equilibrium abundances of ${\rm H_2}$ and associated carbon tracers (${\rm CO}$, ${\rm C}$, and ${\rm C^+}$) within cosmological simulations of galaxy formation. The paper is organised as follows: in Sect.~\ref{sec:methods}, we describe the components of HYACINTH and how it can be incorporated into cosmological simulations. A comparison of our chemical network with two more complex chemical networks and a photon-dominated region code is presented in Sect.~\ref{sec:nl99_g17}. We further compare the chemical evolution from HYACINTH against high-resolution simulations of molecular clouds -- the SILCC-Zoom simulations \citep{silcc-zoom, seifried20} and the \cite{glover11} simulations in  Sect.~\ref{sec:silcc}. Although HYACINTH is primarily designed to be embedded as a sub-grid model in cosmological simulations, as an immediate application, we applied it to a galaxy simulation (from T15) in post-processing and directly compared it with observations related to the abundances of ${\rm H_2}$, ${\rm CO}$, ${\rm C}$, and ${\rm C^+}$ in nearby and high-redshift galaxies. These are discussed in Sect.~\ref{sec:results}. In Sect.~\ref{sec:discussion}, we elaborate on the caveats involved in using HYACINTH as a post-processing tool and compare this with other approaches in the literature. A summary of our findings is presented in Sect.~\ref{sec:summary}.
%--------------------------------------------------------------------

\section{Methods}
\label{sec:methods}
At the core of HYACINTH is a PDF of sub-grid densities and a simplified chemical network. The PDF accounts for the unresolved density structure in simulations, statistically incorporating its impact on the chemistry at resolved scales. The chemical network for hydrogen and carbon chemistry is designed to be efficient and scalable for integration into large-scale galaxy formation simulations. Additionally, HYACINTH can be used as a post-processing tool as well; see Sect.~\ref{sec:results} for a sample application. In the following, we describe the technical specifications of these two components. 

\subsection{The sub-grid density PDF}
\label{sec:2.1} 
\begin{figure}
    \centering    
    \includegraphics[width=0.5\textwidth,trim={0 1cm 0.5cm 0 },clip]{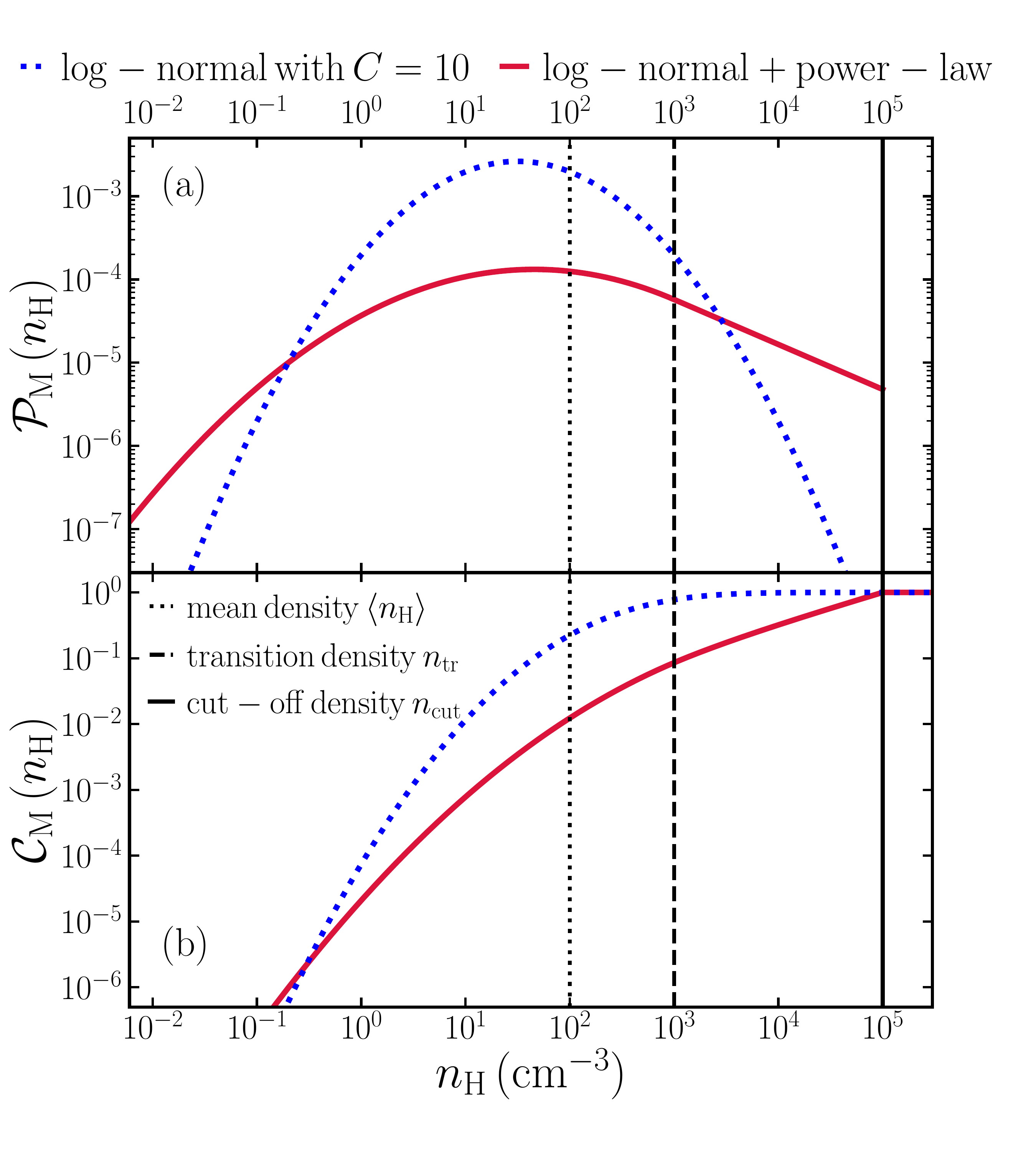}
    \caption{Probability distribution functions (PDFs) and the corresponding cumulative distribution functions (CDFs) used in this study. Panel (a) shows the mass-weighted PDFs in sample simulation cells as a function of the sub-grid density $n_{\rm H}$, where $\mathcal{P}_{\rm M} (n_{\rm H}) \, {\rm d} n_{\rm H}$ denotes the fraction of the total cell mass present at sub-grid densities in the range $[n_{\rm H}, n_{\rm H} +{\rm d} n_{\rm H}]$. The log-normal (Eq.~\ref{eq:log-normal}) and the log-normal+power-law (Eq.~\ref{eq:log-normal+powerlaw}) PDFs are shown in blue and red, respectively. The sample cells have a mean hydrogen density $\langle n_{\rm H} \rangle =100\,\rm cm^{-3}$ (shown by the dotted black line). For the log-normal+power-law PDF, the transition density $n_{\rm tr}$ and the cut-off density $n_{\rm cut}$ are shown by the dashed and solid black lines, respectively. Panel (b) shows the corresponding CDFs, that is, $\mathcal{C}_{\rm M} (n_{\rm H})$ denotes the fraction of the total cell mass present at sub-grid densities below $n_{\rm H}$. For the log-normal+power-law PDF shown here, the power-law tail encloses $\sim 91 \%$ of the total cell mass.}
    \label{fig:pdf_cdf}
\end{figure}
The density structure of molecular clouds is governed by the interplay between turbulence and self-gravity. The PDF of densities is an important statistical property that describes this structure. For instance, the mass-weighted PDF gives the probability that an infinitesimal mass element $\mathrm{d} M$ has a density in the range $[\rho, \rho + \mathrm{d} \rho]$.  This distribution is expected to take a log-normal shape in an isothermal, turbulent medium, not significantly affected by the self-gravity of gas \citep[see e.g.][for a review]{vsemadeni94, passot-vsemadeni98,  mckee07}. Using near-infrared dust extinction mapping of nearby molecular clouds, \cite{kainulainen09} found that the column density PDF in quiescent clouds is very well described by a log-normal. However, in star-forming clouds, they found large deviations from a log-normal: the PDF in these clouds has a log-normal shape only at low column densities and resembles a power law at high column densities. A similar picture is supported by numerical simulations of turbulent and self-gravitating gas \citep[e.g.][]{nordlund-padoan99, klessen00, glover07b, ballesteros-paredes11, ward14}. On scales where the turbulence is supersonic and self-gravity is unimportant, the gas follows a log-normal distribution. In regions where gas is collapsing under its own gravity, the density structure is log-normal at low densities and develops a power-law tail at high densities. Motivated by these findings, we have modified the original PDF of T15 and use a log-normal PDF for turbulence-dominated regions and a log-normal+power-law PDF for gravity-dominated regions. 

The mass-weighted log-normal PDF of sub-grid densities $n_{\rm H}$ is given by (T15)
\begin{equation}
   \mathcal{P}_{\rm M} (n_{\rm H}) \,=\, 
    \frac{1}{\sqrt{2 \pi} \sigma n_{\rm H}} \exp \left [ {-\frac{(\ln{n_{\rm H}}-\mu)^2}{2\sigma^2}} \right] \; ,
    \label{eq:log-normal}
\end{equation}
where $\sigma$ and $\mu$ are parameters that decide the width and the location of the peak of the distribution. It is often convenient to introduce the clumping factor 
\begin{equation}
\label{eq:clumping_factor}
    C=\langle n_{\rm H}^2 \rangle/\langle n_{\rm H} \rangle^2
\end{equation}
that captures the inhomogeneity and the degree of clumpiness in the medium. For a log-normal PDF, it is related to the parameter $\sigma$ above as $C = e^{\sigma^2}$.  We use a constant clumping factor of 10, which has previously been shown to reproduce the observed ${\rm H_2}$ fractions in nearby galaxies \citep[see e.g.][T15]{gnedin09, christensen12}. We note, however, that it is possible to adopt a more sophisticated approach by varying $C$ with the local Mach number $\mathcal{M}$ or the 3D velocity dispersion \citep[see e.g.][]{lupi18}. The parameter $\mu$ is related to the mean density $\langle n_{\rm H} \rangle$ in the region as $\mu = \ln{\langle n_{\rm H} \rangle} + \frac{\sigma^2}{2}$.

For a log-normal+power-law distribution of $n_{\rm H}$, the  PDF takes the form 
\begin{equation}
    \mathcal{P}_{\rm M} (n_{\rm H}) \,=\, 
    \begin{cases}
        \displaystyle\frac{Q_1}{ n_{\rm H}} \exp \left [ {-\frac{(\ln{n_{\rm H}}-\mu_2)^2}{2\sigma_{2}^2}} \right] ,\,& \begin{aligned} \mathrm{if}\ n_{\rm H} \leq n_{\rm tr} \end{aligned} \\[10pt]
        \displaystyle Q_2 \,
        \left ( \frac{n_{\rm H}}{n_{\rm tr}}\right )^{\alpha} ,\, & \begin{aligned} \mathrm{if}\ n_{\rm tr} < n_{\rm H} \leq n_{\rm cut}  \end{aligned} \\[10pt]
        \displaystyle 0 , \, & \begin{aligned} \mathrm{if}\ n_{\rm H} > n_{\rm cut} \end{aligned},
    \end{cases}
    \label{eq:log-normal+powerlaw}
\end{equation}
where $\alpha<0$ is the slope of the power law and $n_{\rm tr}$ is the density at which the power-law tail begins. The parameters $\mu_2$ and $\sigma_2$ characterise the location of the peak and the width of the log-normal part of the PDF. These are calculated, along with constants $Q_1$ and $Q_2$, for a given $\langle n_{\rm H} \rangle$, $\alpha$, and $n_{\rm tr}$ to match the mean density to $\langle n_{\rm H} \rangle$ and ensure the continuity, differentiability, and normalisation of the PDF. Numerical simulations of self-gravitating molecular clouds \citep[e.g.][]{kritsuk11, federrath13} have shown that a power law with an index $\alpha>-1$  provides a good fit to the density distribution in regions undergoing gravitational collapse. \cite{girichidis14} developed an analytical model for the time evolution of the slope of the power-law tail in a spherically collapsing cloud. They found that irrespective of the initial state of the cloud, after one free-fall time, the power-law tail has a universal slope of $-0.54$. Their value is in good agreement with the range of values observed for star-forming clouds in \cite{kainulainen09} and those found by \cite{kritsuk11} in simulations\footnote{Several studies \citep[e.g.][]{kritsuk11} use a volume-weighted density PDF. The power-law index $\alpha_{\rm V}$ of the volume-weighted PDF is related to the power-law index $\alpha_{\rm M}$ of the mass-weighted PDF as $\alpha_{\rm V} + 1 = \alpha_{\rm M}$. Unless otherwise stated, we use $\alpha$ to refer to $\alpha_{\rm M}$.}. Therefore, we adopt $\alpha=-0.54$. Furthermore, we set $n_{\rm tr}$ equal to $ 10$ times the mean density $\langle n_{\rm H} \rangle$ \citep{kritsuk11} and, in order to prevent the integral of the PDF from diverging, we impose a cut-off of $n_{\rm cut} = 1000\, \langle n_{\rm H} \rangle$, above which the PDF is set to zero.

For both PDFs, we do not vary the parameters with the spatial resolution as long as it is larger than the typical scales of density fluctuations (i.e. the scales at which clumping takes place) in molecular clouds. For a detailed discussion on the variability of the clumping factor, we refer the interested readers to \cite{micic12} and \cite{alex_h2}. 

Fig.~\ref{fig:pdf_cdf} shows the distribution of sub-grid densities $n_{\rm H}$ in two sample simulation cells with different PDFs, each with $\langle n_{\rm H} \rangle =100 \, \rm cm^{-3}$. The corresponding cumulative distribution functions (CDF) are plotted in the bottom panel. We can see that the log-normal+power-law distribution spans a broader range of densities as compared to the log-normal. This information is captured by the clumping factor $C$, which for the log-normal+power-law PDF used here is $\sim 300$ compared to $C=10$ for the log-normal.

\subsubsection{Effect of stellar feedback on the PDF}
\label{sec:stellar_feedback}

The internal structure and lifecycle of molecular clouds are affected by star formation and stellar feedback. While supernova (SN) feedback begins to act only a few million years (3-10) after star formation, pre-SN feedback in the form of stellar winds, photoionisation, and radiation pressure, particularly the latter in dense regions, can already start to act once stars form within molecular clouds. Recent observations of molecular clouds in nearby galaxies \citep{hollyhead15, hannon19,  kruijssen19, chevance20a} have found that these effects efficiently disperse the gas within molecular clouds.

A typical molecular-cloud region would cycle through episodes of star formation followed by stellar feedback. We use a log-normal+power-law PDF for gravitationally collapsing regions before star formation. At the onset of star formation in this region, we transition to a log-normal PDF to capture the combined effects of pre-SN and SN feedback. After a period of 40 Myr (\citealt{oey97}; also see \citealt{chevance23} for a recent review of molecular cloud lifetimes), we switch back to the log-normal+power-law PDF. This 40 Myr timescale includes the molecular cloud dispersal time plus the time it takes for the assembly of the next cloud, but still prior to collapse. We note that the exact value of this timescale time will vary as a function of the local density and might be overestimated for some regions. 
  
We note that our model does not explicitly capture the collapse of gas and star formation but only approximates their effect on chemistry by modifying the PDF from a log-normal to a log-normal+power-law form. Hence, these PDFs are a tool to mimic the effects of the `microscopic' (i.e. unresolved) density structure on the `macroscopic' (resolved) chemistry at different stages in the lifecycle of a molecular cloud.

%----------------------------------------------
%----------------------------------------------
%----------------------------------------------
\subsection{The chemical network}
\label{sec:2.2}

We use a simplified version of the \citet[][hereafter NL99]{NL99} chemical network with some modifications (described in the following subsections) from the recent work of \citet[][hereafter G17]{GOW17}. Our simplifications reduce the number of chemical species and reactions that we follow to retain only the dominant formation and destruction channels for ${\rm H_2}$, ${\rm CO}$, ${\rm C}$, and ${\rm C^+}$ under the physical conditions prevalent in molecular clouds (see Table~\ref{tab:chem_reactions} in appendix~\ref{appendixA}). This simplifies the NL99 network for easier integration into cosmological simulations, allowing on-the-fly computation of chemical abundances without a significant increase in the computational overhead of these simulations. 

Numerical simulations of molecular clouds \citep[see e.g.][]{glover07b, hu21} show that at the densities relevant for ${\rm H_2}$ formation in molecular clouds, the temperatures are mostly $\lesssim 200 \, \rm K$. At these temperatures, the contribution of ionised hydrogen ($\rm H^+$) to the total hydrogen mass is expected to be negligible. Therefore, we make a further simplifying assumption that all hydrogen is either atomic or molecular. Hence, in practice, we solve the system of rate equations for only three species, namely ${\rm H_2}$, ${\rm CO}$, and ${\rm C^+}$. The electron abundance follows from change conservation (i.e. $f_{\rm e^-} = f_{{\rm C^+}}$). We compute the abundances of ${\rm H}$ and ${\rm C}$ based on the conservation of hydrogen and carbon nuclei, respectively. We assume that the total gas-phase elemental abundances of carbon and oxygen are proportional to the gas metallicity $Z$, i.e. $f_{\rm C, tot} = 1.41 \times 10^{-4} \, (Z/\rm Z_{\odot})$ and $f_{\rm O, tot} = 3.16 \times 10^{-4} \, (Z/\rm Z_{\odot})$ \citep{savage96}.

%-----------------------------------------------
\subsubsection{${\rm H_2}$ chemistry}
\label{sec:h2_chem}
We consider the following formation and destruction channels for ${\rm H_2}$ -- 
\begin{enumerate}
    \item ${\rm H_2}$ formation on dust grains
    $$ {\rm H} + {\rm H} + \rm{grain} \rightarrow {\rm H_2} + \rm{grain} \, ; $$
    \item radiative recombination of ${\rm H_3^+}$ with an electron
    $$ {\rm H_3^+} + {\rm e^-} \rightarrow {\rm H_2} + {\rm H}  \, ; $$ 
    \item  collisional reactions of ${\rm H_3^+}$ with atomic carbon and oxygen
    $$ {\rm H_3^+} + {\rm C} \rightarrow {\rm CH_{\rm x}} + {\rm H_2}  \, , $$
    $$ {\rm H_3^+} + {\rm O} \rightarrow {\rm OH_{\rm x}} + {\rm H_2}  \, ; $$    
    \item photodissociation of ${\rm H_2}$
    $$ {\rm H_2} + \gamma \rightarrow 2 {\rm H} \, ; $$
    \item ionisation of ${\rm H_2}$ by cosmic rays
    $$ {\rm H_2} + \rm{CR} \rightarrow {\rm H_2^+} + {\rm e^-} \, ; $$
    \item collisional reaction of ${\rm H_2}$ with ${\rm C^+}$
    $$ {\rm H_2} + {\rm C^+} + {\rm e^-} \rightarrow {\rm CH_{\rm x}} + {\rm H}  \, , $$    
    $$ {\rm H_2} + {\rm C^+} + {\rm e^-} \rightarrow {\rm C} + 2 {\rm H}  \, . $$    

\end{enumerate}
The rate of ${\rm H_2}$ formation on dust grains depends on the dust abundance. We adopt a metallicity-dependent dust-to-gas mass ratio (DTG) based on observational measurements and theoretical predictions of the DTG as a function of the gas-phase metallicity in galaxies at redshifts $0<z\lesssim5$ (\citealt{celine_review} and \citealt{popping22}). The DTG in our model is given as:
\begin{equation}
    \rm log_{10}\,(DTG) = 1.3 \, log_{10}\,(Z/Z_{\odot}) - 2.02 \, ,
\end{equation}
where $Z/\rm Z_{\odot}$ is the gas-phase metallicity in solar units. We use $\rm Z_{\odot}=0.02$ \citep{karakas10}. The above non-linear dependence of the DTG on gas metallicity reflects a variable dust-to-metals (DTM) ratio (i.e. the fraction of metals locked up in dust grains). Based on a compilation of absorption-line studies of high-redshift objects, \cite{popping22} found that the DTG does not show significant evolution for $0<z<5$.  Hence, we further assume that the same relation holds at all redshifts.

For the collisional reaction between ${\rm H_2}$ and ${\rm C^+}$, NL99 only considers the first outcome. However, it has been shown by \cite{wakelam10} that the reaction between ${\rm C^+}$ and ${\rm H_2}$ gives ${\rm CH_{\rm x}} + {\rm H}$ only $70\%$ of the times. In the remaining $30 \%$ of the cases, ${\rm C} + 2 {\rm H}$ are formed instead. This has important consequences for the relative abundances of ${\rm C}$ and ${\rm CO}$. The ${\rm CH_{\rm x}}$ formed in the first outcome can react with an ${\rm O}$ atom to form ${\rm CO}$ or could photodissociate into ${\rm C}$ and ${\rm H}$ atoms, whereas, the second outcome acts as an additional formation channel for ${\rm C}$ in our network as well as in G17. The dissociation of ${\rm H_2}$ is carried out by photons in two narrow bands of energies in the range $11.2-13.6\, \rm eV$ ($\lambda = 912-1108 \, \AA$), called Lyman-Werner photons. We do not explicitly include three-body interactions in our network as these are inefficient at most ISM densities resolved in cosmological simulations. However, three-body reactions are the main mechanism for ${\rm H_2}$ formation at high redshifts $z\gtrsim 12$ \citep[see e.g.][]{christensen12, lenoble24}.

\subsubsection{${\rm H_3^+}$ chemistry} 
\label{sec:h3+_chem}
The ionisation of ${\rm H_2}$ by cosmic rays produces the ${\rm H_2^+}$ ion that quickly reacts with an ${\rm H}$ atom to form ${\rm H_3^+}$. ${\rm H_3^+}$ can be destroyed by reactions with ${\rm e^-}$, ${\rm C}$, and ${\rm O}$. Because of its high reactivity \citep{oka06}, we assume a local (i.e. at each sub-grid density) equilibrium between its formation and destruction channels (see Eqs.~\ref{eq:h3p1}-~\ref{eq:h3p2} in Appendix~\ref{appendixA}). We note that NL99 also include the reaction of ${\rm H_3^+}$ with ${\rm CO}$, which is excluded from our chemical network to limit the number of chemical species and reactions. Another difference with respect to NL99 is that, following G17, we consider two outcomes for the recombination of ${\rm H_3^+}$ with ${\rm e^-}$: a) ${\rm H_2} + {\rm H}$ b) 3 ${\rm H}$. Of these, only the first one is included in NL99.

\subsubsection{CO chemistry}
The formation of ${\rm CO}$ proceeds via the reaction of ${\rm CH_{\rm x}}$(${\rm OH_{\rm x}}$) with ${\rm O}\,({\rm C})$ which is formed by the reaction of ${\rm H_3^+}$ with ${\rm C} \, ({\rm O})$ (see reactions 3-6 in Table~\ref{tab:chem_reactions}). ${\rm CH_{\rm x}}$ can additionally be formed by the collisional reaction between ${\rm C^+}$ and ${\rm H_2}$. The destruction channels for ${\rm CO}$ are dissociation into ${\rm C}$ and ${\rm O}$ by UV photons in the $912-1100 \, \AA$ band and cosmic rays. The rate of photodissociation drops off exponentially with the increasing column density of ${\rm H_2}$, ${\rm CO}$, and dust as a result of the shielding effect of these species on the impinging UV radiation deep inside molecular clouds (see Sect.~\ref{sec:shielding_fn} for the expression of the shielding functions).

\subsubsection{Grain-assisted recombination of ${\rm C^+}$}
\label{sec:grain-assisted_recom}

Following \citet[][hereafter GC12]{glover12b} and G17, we include grain-assisted recombination of ${\rm C^+}$ in our network in addition to its radiative recombination. This is the main channel for ${\rm C^+}$ recombination at solar metallicity (G17), although its importance at sub-solar metallicities would depend on the relative amount of dust to gas in the ISM (the dust-to-gas ratio, see Sect.~\ref{sec:h2_chem}). Moreover, for several ions, including ${\rm C^+}$, the recombination rate on dust grains is often higher than the direct radiative recombination rate, especially in star-forming regions where dust, particularly in the form of polycyclic aromatic hydrocarbons (PAHs) is highly prevalent and frequently collides with ions leading to their recombinations. In a previous study, GC12 stressed that this reaction is particularly effective when the ratio of the UV field strength to the mean density is very small. As a result of including this additional destruction (formation) channel for ${\rm C^+}$ (${\rm C}$), we expect the ${\rm C}/{\rm C^+}$ ratio predicted by our network (as well as in G17) to be significantly higher than predicted by NL99 (see Sect.~\ref{sec:nl99_g17}).

\subsubsection{Cosmic rays}
\label{sec:cr}
Cosmic rays (CRs) with energies $\lesssim 0.1 \, \rm GeV$ play a critical role in initiating ion-ion chemistry deep inside dense molecular-cloud regions that are well-shielded from UV radiation (see e.g. \citealt{padovani09}). These reactions become particularly relevant at high redshifts where the cosmic ray ionisation rate per {\rm H} atom (CRIR, $\zeta_{\rm H} $) is expected to be higher than the canonical MW value of $3 \times 10^{-17} \, \rm s^{-1}$ because of higher star formation rates \citep[see e.g.][for CRIR estimates at high redshifts]{muller16, indriolo18}. We include the ionisation of ${\rm H_2}$, the ionisation of ${\rm C}$, and the dissociation of ${\rm CO}$ by CRs in our network while NL99 only includes the first one.

Based on absorption studies of $\rm HD$ in ${\rm H_2}$-bearing damped Lyman$-\alpha$ systems, \cite{kosenko21} found that the CRIR scales quadratically with the UV field intensity relative to the Draine field\footnote{Often the UV field strength is expressed in terms of the Habing Field \citep{habing68} and the parameter $G_0$ captures the ratio between the energy density of a given UV field and the energy density measured in the solar neighbourhood by \cite{habing68}. The Draine field has $G_0=1.7$.} \citep[$\chi$,][] {draine78}. Therefore, as a default choice, we adopt the following relation between $\zeta_{\rm H} $ and $\chi$: 
\begin{equation}\label{eq:crir}
    \frac{\zeta_{\rm H}}{\zeta_{\rm H, MW}} =\left (\frac{\chi}{\chi_{\rm MW}} \right) ^2 \, , 
\end{equation}
but also consider alternative options in Sect.~\ref{sec:results} and Appendix~\ref{appendixD}. We use $\zeta_{\rm H, MW}=3\times10^{-17}\, \rm s^{-1}$ and $\chi_{\rm MW}=1.0$ for MW. We also impose upper and a lower bounds on the CRIR of $3\times 10^{-14}\,\rm s^{-1}$ and $10^{-18}\,\rm s^{-1}$, respectively, to avoid unreasonable CRIRs. The effect of this upper limit is investigated in Sect.~\ref{sec:results} and Appendix~\ref{appendixD}. We note that the $\chi$ in Eq. (\ref{eq:crir}) is measured in the FUV band ($\lambda = 912-2070 \, \AA$), while HYACINTH requires the UV flux in the Lyman -Werner band ($\lambda = 912-1080 \, \AA$) in Habing units as an input. In the solar neighbourhood, the mean energy densities in the two bands are related as: $U_{\rm LW} \, / \, U_{\rm FUV} \sim 1.1$ \citep{parravano03}.

\subsubsection{The temperature-density relation}
\label{sec:t-rho}
As some chemical reactions in our network have temperature-dependent rate coefficients (see Table~\ref{tab:chem_reactions}), we need to associate a temperature with each sub-grid density in the PDF. For this, we use a metallicity-dependent temperature-density relation obtained from simulations of the ISM \citep{hu21}. These simulations self-consistently include time-dependent ${\rm H_2}$ chemistry and cooling, star formation, and feedback from photoionisation and supernovae. They adopt a linear relationship between the CRIR and the UV intensity (actually both quantities scale with the star-formation-rate surface density) which differs from our quadratic scaling and thus leads to a small inconsistency in our assumptions for dense, well-shielded gas, where CRs are an important heating source. In Sect.~\ref{sec:results}, we compare the chemical abundances obtained when using the $\zeta_{{\rm H}}-\chi$ relation from \cite{hu21} against those obtained from Eq.~(\ref{eq:crir}).

\subsubsection{Shielding functions}\label{sec:shielding_fn}
Dense, optically thick gas can shield itself against penetrating UV radiation because of the high column densities of ${\rm H_2}$, ${\rm CO}$, ${\rm C}$, and dust. We account for this by modifying the reaction rates of photoionisation and photodissociation reactions by appropriate shielding functions. The (self-) shielding of ${\rm H_2}$ is given by \citep{draine_bertoldi}:
\begin{eqnarray}
    \begin{aligned}
    &f_{s, {\rm H_2}} \left( N_{\rm H_2} \right)=\frac{0.965}{\left(1+x / b_{5}\right)^{2}} \\
    &+\frac{0.035}{(1+x)^{0.5}} \exp \left[-8.5 \times 10^{-4}(1+x)^{0.5}\right] \, ,
\end{aligned} 
\end{eqnarray}
where $x=\frac{N_{\rm H_2}}{5\times10^{14}\,\rm cm^{-2}}$, $N_{\rm H_2}$ is the column density of ${\rm H_2}$ and $b_5$ is the velocity dispersion of gas in $\rm km \, s^{-1}$. Following \cite{sternberg14} and T15,  we use a constant $b_5=2$ throughout. However, as noted in G17, the ${\rm H_2}$ fraction $f_{\rm H_2}$ is insensitive to the value of $b_5$ for $f_{\rm H_2} \gtrsim 0.1$. 

The ${\rm CO}$ shielding function $f_{s,{\rm CO} }\left( N_{\rm CO}, N_{\rm H_2} \right)$ accounts for both ${\rm CO}$ self-shielding and the shielding of ${\rm CO}$ by ${\rm H_2}$ and is calculated by interpolating over $N_{\rm CO}$ and $N_{\rm H_2}$ from Table 5 in \citet{visser09}. 

The shielding function for ${\rm C}$ is given by \citep{tielens_hollenback85}
\begin{equation}
    f_{\mathrm{s}, \mathrm{C}}\left(N_{\mathrm{C}}, N_{\mathrm{H}_{2}}\right)=\exp \left(-\tau_{\mathrm{C}}\right) f_{\mathrm{s}, \mathrm{C}}\left(\mathrm{H}_{2}\right) \, ,
\end{equation}
where 
\begin{equation}
\begin{aligned}
\tau_{{\rm C}} = 1.6 \times 10^{-17}\, \displaystyle\frac{N_{\rm C}}{\rm cm^{-2}} \, ,\\
f_{s,{\rm C}}\left( {\rm H_2} \right) = \displaystyle\frac{\exp{(-r_{{\rm H_2}})}}{1+r_{{\rm H_2}}} \, , \\
r_{{\rm H_2}} = 2.8 \times 10^{-22} \displaystyle\frac{N_{\rm H_2}}{\rm cm^{-2}} \, ,
\end{aligned}
\end{equation}
where $N_{\rm C}$ and $N_{\rm H_2}$ are the column densities of atomic carbon and ${\rm H_2}$, respectively.

In addition to self-shielding and ${\rm H_2}-$shielding, all species are also shielded against UV radiation by dust grains; the relevant shielding function is given by 
\begin{equation} \label{eq:dust_shielding}
    f_{\rm dust} = \exp{\left( -\gamma A_V \right)} \, ,
\end{equation}
where $\gamma$ is different for each species and listed in Table~\ref{tab:chem_reactions} in Appendix~\ref{appendixA}. The visual extinction $A_V$ is related to the total column density of hydrogen nuclei $N_{\rm H}=N_{\rm H \, \textsc{i}}+2N_{\rm H_2}$ along the line of sight as
\begin{equation}\label{eq:Av}
    A_V = \frac{N_{\rm H} \,Z_{\rm d}}{1.87\times10^{21}\,\rm cm^{-2}} \, .
\end{equation}

\vspace{0.5cm}
In a simulation, recording the ${\rm H_2}$ fraction for every sub-grid density within a grid cell at each timestep is computationally expensive and impractical. Therefore, we follow the approach of T15 for distributing the available ${\rm H_2}$ to the different sub-grid densities. We assume a sharp transition from atomic to molecular hydrogen at the sub-grid density $n_{\rm H} = n_{\rm crit, H_2}$, that is $f_{\rm H_2}(n_{\rm H} < n_{\rm crit, H_2}) = 0$ and $ f_{\rm H_2}(n_{\rm H} \geq n_{\rm crit, H_2}) = 1$. This sharp transition has been observed in various numerical studies \citep{glover07a, dobbs08, kmt08, kmt09, gnedin09}, and occurs at densities where ${\rm H_2}$ becomes self-shielding \citep{dobbs14}. Similarly, we assume that carbon transitions from ionic to atomic form above $n_{\rm crit, \rm C\,\rm \textsc{i}}$ and becomes fully molecular at $n_{\rm crit, \rm CO}$. In a given region, these critical densities depend on the density of total hydrogen and the abundance of the different species involved in the transition (see Appendix~\ref{appendixC} for calculation of $n_{\rm crit, H_2}$ for the ${\rm H}\,\rm \textsc{i} - {\rm H_2} $ transition).

In practice, our sub-grid model requires six input parameters --  the average density of hydrogen nuclei $\langle n_{\rm H} \rangle$, the gas-phase metallicity $Z$, the UV flux in Lyman-Werner bands in Habing units $G_0$, the characteristic length scale $\Delta x$ of a resolution element (e.g. cell size or smoothing length), the density PDF $\mathcal{P}_{\rm M}$, and the time $\Delta t$ over which the chemical abundances are to be evolved. When embedded as a sub-grid model in a simulation, these parameters can be obtained directly from the simulation or calculated in post-processing (e.g. the UV flux). Depending on the value of Z, the temperature-density relation assigns a (sub-grid) temperature to each sub-grid density in the PDF. The chemical network then solves the rate equations for each sub-grid density and integrates over the PDF to obtain the average abundances of ${\rm H_2}$, ${\rm CO}$, ${\rm C}$, and ${\rm C^+}$ with respect to the total hydrogen within the region. 

%--------------------------------------------------------------------

\section{Comparison of chemical abundances}
In this section, we compare different aspects of HYACINTH to previous approaches in the literature. First, we focus on the chemical network and contrast its predictions to those from the NL99 and G17 networks. Subsequently, we compare our full implementation of the sub-grid model (including the density PDF) with the output of high-resolution simulations of individual molecular clouds \citep{silcc-zoom, seifried20, glover11}.

\subsection{Comparison with NL99 and G17}
\label{sec:nl99_g17}
\begin{figure*}
    \centering   
    \includegraphics[width=0.98\textwidth, trim={0 4cm 0 4cm },clip]{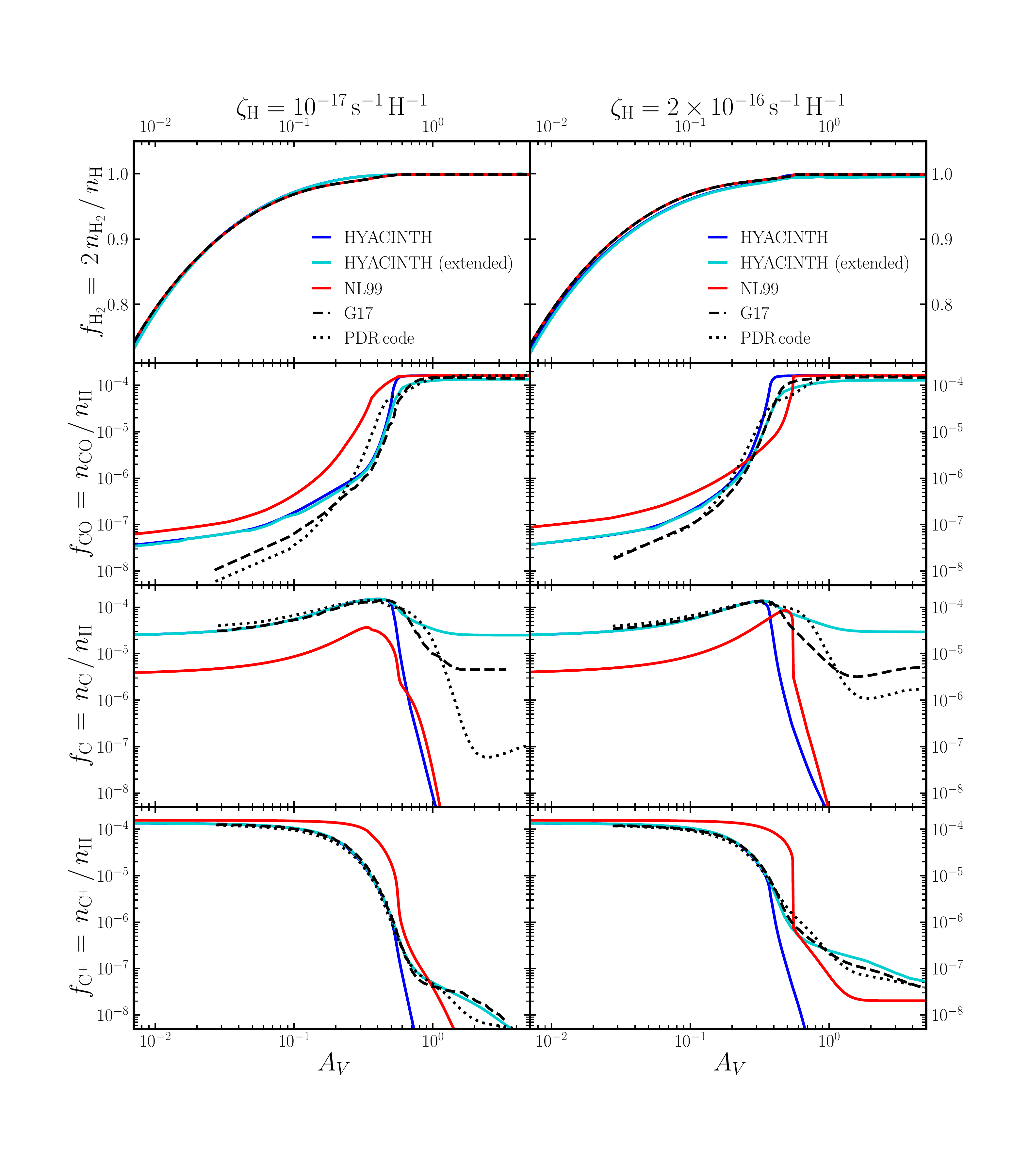}
    \caption{Comparison of the chemical network in HYACINTH with NL99 and G17 networks. The abundances of  ${\rm H_2}$, ${\rm CO}$, ${\rm C}$, and ${\rm C^+}$ as a function of the visual extinction $A_V$ in a semi-infinite plane-parallel slab are shown in different panels. The left column shows the abundances for a CRIR ($\zeta_{\rm H}$) of $10^{-17} \, \rm s^{-1} \, {\rm H}^{-1}$ while the right one has $\zeta_{\rm H} = 2 \times10^{-16} \, \rm s^{-1} \, {\rm H}^{-1}$.  The blue, turquoise, and red lines represent the results from HYACINTH, extended HYACINTH, and NL99 respectively. Here extended HYACINTH refers to the HYACINTH network with additional chemical reactions for $\rm He^+$ and $\rm HCO^+$ that are not part of standard HYACINTH (see text for more details). The dashed and dotted black lines show, respectively, the abundances from the chemical network and the PDR code in G17. The slab has a uniform hydrogen density $n_{\rm H} = 1000 \, \rm cm^{-3}$, solar metallicity and solar dust abundance, and is illuminated from one side by a UV field of strength $\chi=1$. }
    \label{fig:nl99_g17}
\end{figure*}
 
In this section, we compare our chemical network predictions to Figure 2(b) and Figure 3(b) in G17. The setup involves a one-dimensional semi-infinite slab with a uniform hydrogen density $n_{\rm H} = 1000 \, \rm cm^{-3}$, a solar metallicity ($Z=\rm Z_{\odot}$), a solar dust-to-gas ratio of 0.01, and fixed gas and dust temperatures of $20\, \rm K$ and $10 \, \rm K$, respectively. The slab is irradiated from one side by a UV field of strength $1$ in Draine units (i.e. $\chi=1$). We test for two different values of the CRIR (same as in G17): $1 \times 10^{-17} \, \rm s^{-1} \, \rm H^{-1}$ (left column of Fig.~\ref{fig:nl99_g17}) and  $2 \times 10^{-16} \, \rm s^{-1} \, \rm H^{-1}$ (right column of Fig.~\ref{fig:nl99_g17}). For this comparison, we assume the same elemental abundances for carbon and oxygen as in G17, that is, $f_{\rm C, tot} = 1.6 \times 10^{-4}  \,(Z/\rm Z_{\odot})$ and $f_{\rm O, tot} = 3.2 \times 10^{-4} \, (Z/\rm Z_{\odot})$, throughout Sect.~\ref{sec:nl99_g17}. 

For a fair comparison with the G17 results, we calculate the shielding to the incident radiation field using their approximation for an isotropic radiation field \citep[as described in section 2.3 of G17 and originally used by ][]{wolfire10}. Briefly, it approximates an isotropic radiation field with a unidirectional field incident at an angle of $60^\circ$ with the normal to the slab. Thus, for an incident radiation field of strength $\chi$, the effective radiation field at a (perpendicular) depth $L$ into the slab can be expressed as  
$$\chi_{\rm eff} = (\chi/2) \, f_{\rm shield}(2 L) \, ,$$
where $f_{\rm shield}(L)$ represents the shielding function at a depth $L$ into the slab and is different for each chemical species (see Sect.~\ref{sec:shielding_fn}). For this uniform-density slab, the column density of hydrogen, $N_{\rm H}$, at depth $L$ can be written as  $N_{\rm H}(L) = n_{\rm H} \, L$. The slab is divided into $1000$ layers with $N_{\rm H}$ values spaced logarithmically in the range $N_{\rm H} = 10^{17} \, \rm cm^{-2} - 10^{22} \, \rm cm^{-2} $ or equivalently, visual extinction $A_V$ (Eq.~\ref{eq:Av}) in the range $5.35 \times 10^{-5} - 5.35$. For each layer, the NL99 and HYACINTH networks are evolved until equilibrium.  

To put our results in context, we compare them to the output of more complex models. First, we consider the extended HYACINTH chemical network\footnote{The extra chemical reactions treated in extended HYACINTH are listed at the end of  Table~\ref{tab:chem_reactions} (reactions 20-29).} that includes the non-equilibrium treatment of two additional species, namely, $\rm He^+$ and $\rm HCO^+$. These species, particularly $\rm He^+$, serve as the main destruction agents for $\rm CO$ in dense, shielded regions.  Moreover, we present the output of the photon-dominated region (PDR) code used in G17 which tracks the abundances of 74 species accounting for 322 chemical reactions and includes a more sophisticated treatment of radiative transfer. This PDR code is derived from \cite{tielens_hollenback85} and updated by \cite{wolfire10}, \cite{hollenbach2012}, and \cite{neufeld16}. 

Fig.~\ref{fig:nl99_g17} shows the equilibrium abundances of the chemical species as a function of $A_V$. The $f_{\rm H_2}$ versus $A_V$ from our network is in excellent agreement with that from NL99, G17 and the PDR code. In contrast, there are noticeable differences in the ${\rm C^+} \rightarrow {\rm C}$ and ${\rm C} \rightarrow {\rm CO}$ transitions in the different networks. For both values of the CRIR, at low $A_V$, the ${\rm C^+}$ abundance in HYACINTH (both standard and extended), G17 and the PDR code is slightly lower than NL99, accompanied by a higher ${\rm C}$ abundance for these networks with respect to NL99. In HYACINTH, this shift results primarily from the inclusion of grain-assisted recombination of ${\rm C^+}$ (see Sect.~\ref{sec:grain-assisted_recom}) and an additional outcome for the ${\rm C^+} +{\rm H_2}$ reaction (see Sect.~\ref{sec:h2_chem}).  Together, these lead to almost an order of magnitude increase in the abundance of atomic carbon at $A_V < 0.5$ in HYACINTH compared to NL99. GC12 observed a similar trend when comparing their networks with and without this reaction. Previous studies have reported that the chemical networks employed in PDR codes tend to produce an elevated atomic carbon abundance with respect to the atomic carbon abundance measured in MW (interstellar) clouds \citep{sofia04, sheffer08, wolfire08, bfj10}. This has been long-standing problem for several chemical networks \citep[see e.g.][]{bfj10, liszt11, GOW17}.

For $\zeta_{\rm H} = 10^{-17} \, \rm s^{-1} \, H^{-1}$, the ${\rm C} \rightarrow {\rm CO}$ transition in HYACINTH (both standard and extended), G17 and the PDR code occurs at a slightly higher $A_V$ than in NL99. Conversely, for $\zeta_{\rm H} = 2 \times 10^{-16} \, \rm s^{-1} \, H^{-1}$, this transition occurs at a slightly lower $A_V$ in all other approaches compared to NL99.  However, this does not have any practical implications for the modelling of ${\rm CO}$ chemistry in galaxy simulations as we expect the bulk of the ${\rm CO}$ mass in a molecular cloud or a galaxy to be present at high $A_V$ ($\gtrsim 2$). 

A noticeable difference between our network and G17 is that at high $A_V$ ($\gtrsim 1$), all carbon in our network is in the form of ${\rm CO}$, while G17 predicts $\sim 3-10 \%$ of the carbon to be in atomic form. The PDR code predicts that $\lesssim  3 \%$ of the total carbon is in atomic form at $A_V \gtrsim 1$ for both values of the CRIR. In contrast, extended HYACINTH predicts an even higher abundance of atomic carbon than G17 at $A_V \gtrsim 1$ and consequently a lower $\rm CO$ abundance. Therefore, it is evident that the varying complexities of the different approaches result in significant differences in the atomic carbon abundance at $A_V \geq 1$. However, since the bulk of the atomic carbon mass in all networks is present at intermediate $A_V$ ($0.1 \lesssim A_V \lesssim 1$), the contribution of the atomic carbon in G17 and the PDR code at $A_V \geq 1$ to the total atomic carbon mass is $\lesssim 2 \% $ and is therefore, not significant. 

Overall, both the standard and extended HYACINTH networks show a good agreement with NL99, G17, and the PDR code in the respective $A_V$ range where each of the carbon species dominates. For $\rm H_2$, the agreement is excellent for all $A_V$, showing that hydrogen chemistry is not sensitive to the exact treatment of carbon chemistry. There are, however, noticeable differences between standard and extended HYACINTH for the carbon-based species. For instance, there is a significant difference between the abundances of $\rm C$ and $\rm CO$ at $A_V \gtrsim 0.5$. Extended HYACINTH shows a smoother ${\rm C} \rightarrow {\rm CO}$ transition as compared to standard HYACINTH. This transition closely matches that in G17. Nevertheless, at $A_V \gtrsim 2$, the $\rm CO$ abundance in extended HYACINTH is only marginally different from that in the standard one -- roughly $16 \% \, (20\%)$ less for $\zeta_{\rm H} = 10^{-17} \, \rm s^{-1} \, H^{-1} \, (2 \times 10^{-16} \, \rm s^{-1} \, H^{-1})$. This shows that standard HYACINTH provides robust CO abundances with respect to extended HYACINTH, while requiring approximately 3.3 times less computational time. Therefore, in what follows, we only consider the standard HYACINTH network.

%--------------------------------------------
\subsection{Comparison with molecular-cloud simulations}
\label{sec:silcc}
Now we shift our focus to assessing the performance of the chemical network in conjunction with the sub-grid density PDF and compare against high-resolution simulations of individual molecular clouds -- the SILCC-Zoom simulations \citep{silcc-zoom, seifried20} with solar metallicity ($Z= \rm Z_{\odot}$, Sect.~\ref{sec:silcc-zoom}) and those from \citet[][hereafter GML11]{glover11} with $Z=0.1 \, \rm Z_{\odot}$ (Sect.~\ref{sec:gml11}). We note that neither the SILCC-Zoom nor the GML11 runs account for star formation and stellar feedback.

\subsubsection{SILCC-Zoom simulations}
\label{sec:silcc-zoom}
\begin{figure*}
    \centering    
    \includegraphics[width=0.95\textwidth, trim={0 3cm 2cm 1cm },clip]{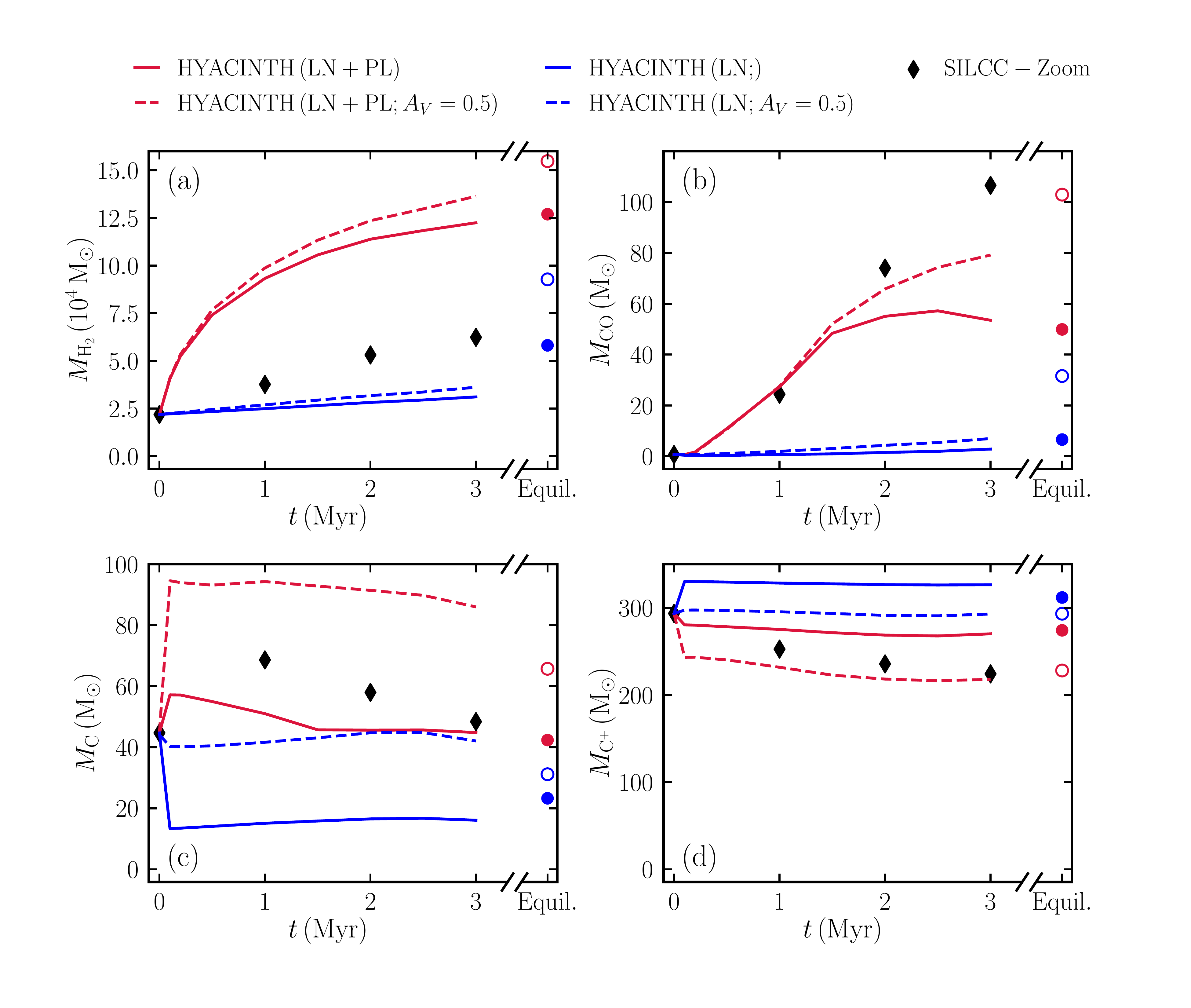}
    \caption{Time evolution of the total mass of different chemical species contained in a $(175 \, \rm pc)^3$ region encompassing the molecular cloud MC1-HD from the SILCC-Zoom simulations (black diamonds).
    The curves show the corresponding results from four low-resolution hydrodynamic simulations that use HYACINTH as a sub-grid model for the chemical evolution. The colour coding distinguishes runs obtained assuming either the LN PDF (blue) or the LN+PL PDF (red). The line style indicates the strength of the assumed uniform UV ISRF: $G_0=1.7$ in Habing units with no attenuation (solid) or with an effective visual extinction of $A_V=0.5$ (dashed). The circles on the right-hand-side of the panels show the masses derived by assuming that the final snapshot of each simulation is in chemical equilibrium. Filled and empty symbols refer to the zero attenuation and $A_V=0.5$ runs, respectively. 
    }
    \label{fig:new_mock_silcc}
\end{figure*}

\begin{figure*}
    \centering    
    \includegraphics[width=0.9\textwidth, trim={0 5cm 0 1cm },clip]{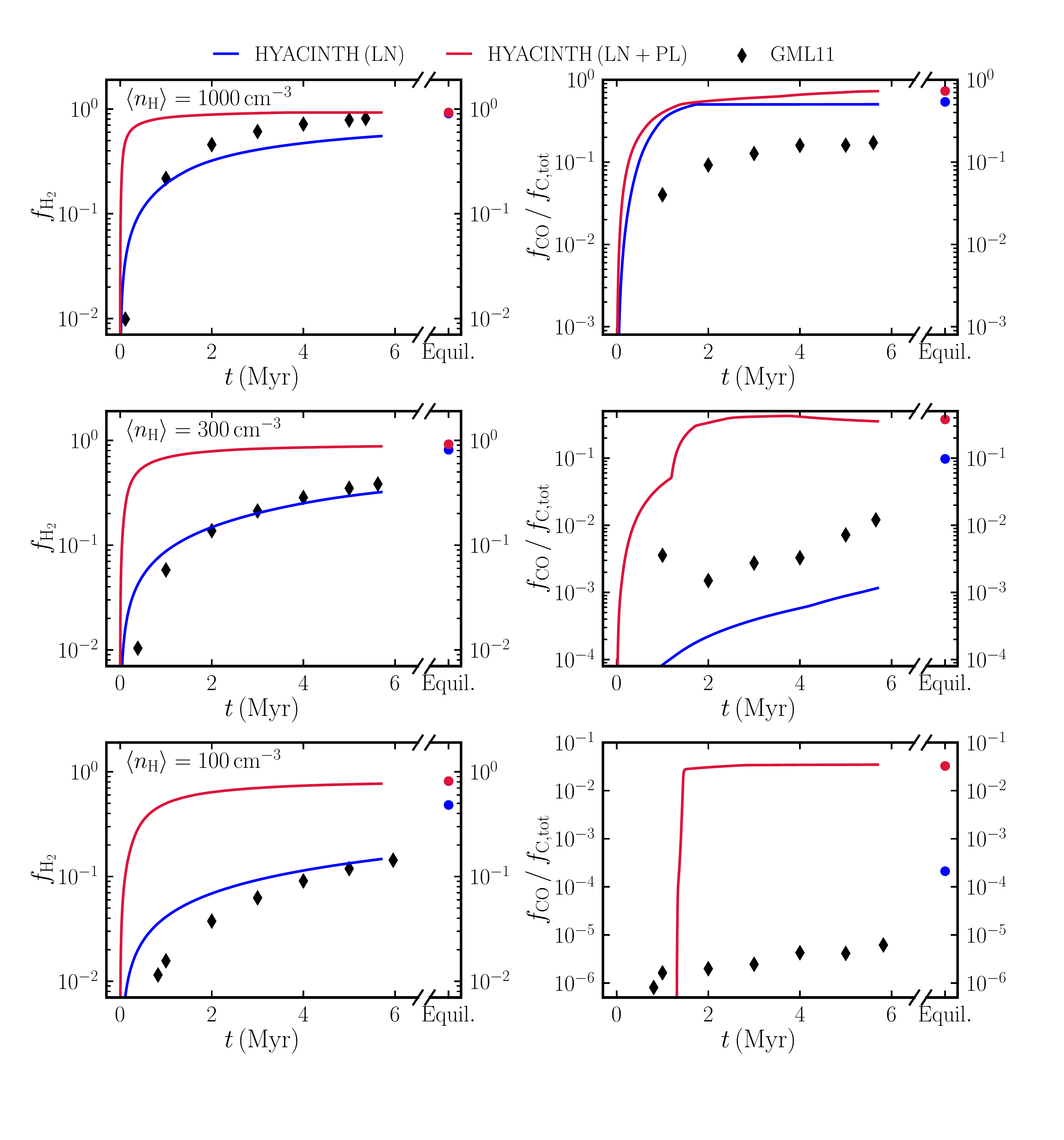}
    \caption{Comparison of the chemical abundances from HYACINTH against those from the $Z = 0.1 \, \rm Z_{\odot}$ runs from GML11 as a function of time. For $\langle n_{\rm H} \rangle = 100 \, \rm cm^{-3}$, very little CO is formed when using the LN PDF (not visible in the bottom-right panel). The circles on the right-hand side denote the equilibrium abundances.}
    \label{fig:mock}
\end{figure*}
%%%%%%%%%%%%%%%%%%%%%%%%%%%%%%%%%%%%%%%%%%%%%%%%%%%%%%%%%%%%%%%%%%%

SILCC-Zoom are adaptive mesh refinement (AMR) simulations that follow the formation of two molecular clouds extracted from the parent SILCC simulations \citep{silcc-i} with the zoom-in technique. They reach a minimum cell size of $0.06\, \rm pc$ in the densest regions. The original SILCC and SILCC-Zoom simulations were run with a chemical network based on \cite{glover07a} and \cite{glover07b} for hydrogen chemistry and \cite{NL97} for ${\rm CO}$ chemistry. However, the SILCC-Zoom simulations used in this comparison \citep{seifried20} were performed with the NL99 network for ${\rm CO}$ chemistry. 

Our goal is to compare the chemical evolution from HYACINTH (employed as a sub-grid model within a hydro simulation) with that from SILCC-Zoom. For this, we build a low-resolution copy of one of the SILCC-Zoom clouds (MC1-HD) by running a simulation with our modified version of the \textsc{Ramses} code \citep{ramses} that uses HYACINTH for evolving the time-dependent chemistry. We consider a cubic box with a side length of $175 \,\mathrm{pc}$ split in cells with a size of $25 \,\mathrm{pc}$ (comparable to the spatial resolution we aim to achieve in our future cosmological simulations). We set the initial conditions (gas properties and chemical abundances) by coarse graining the SILCC-Zoom snapshot after a simulated time of $1 \, \rm Myr$, when all the mesh refinements have been performed, so as to exclude any possible numerical effects of variable spatial resolution on chemical abundances. We then follow the evolution of the system for $3 \, \rm Myr$, that is the full duration of the SILCC-Zoom runs. Beyond self-gravity, our simulation accounts for an external gravitational potential as described in sections 3.1 and 3.2 of \cite{silcc-i}. The DTG is set to $0.01$ and the elemental abundances of carbon and oxygen are set to $f_{\rm C, tot} = 1.41 \times 10^{-4}$ and $f_{\rm O, tot} = 3.16 \times 10^{-4}$. The cells are  irradiated with an ISRF of strength $G_0=1.7$ in Habing units and a CRIR of $\zeta_{\rm H}=1.3 \times 10^{-17}\, \rm s^{-1} \, {\rm H}^{-1}$. 

Before proceeding, two comments are in order. First, the lack of stellar feedback in the SILCC-Zoom simulations leads to the formation of very high-density clumps. Consequently, the density PDFs within the 25 pc cells vary over time and deviate significantly from the analytical forms we adopt in HYACINTH (see also Figure 20 in \citealt{silcc-i} and Figure 3 in \citealt{buck22}). For instance, the clumping factor in these cells spans a wide range of values and is often much higher than the values of 10 and 300 that we adopt for the log-normal (hereafter LN) and log-normal+power-law (hereafter LN+PL) PDFs, respectively. Second, our simulation does not include radiative transfer of UV radiation. Consequently, the local ISRF is not attenuated due to the shielding from surrounding gas and dust as done in the SILCC-Zoom simulations. To quantify the impact of this effect, we perform a second simulation adopting a uniform effective visual extinction $A_V = 0.5$ in the entire simulation volume.

Fig.~\ref{fig:new_mock_silcc} shows the time evolution of the masses of different chemical species in our original ($A_V=0$, solid lines) and UV-attenuated ($A_V=0.5$, dashed lines) simulations compared to SILCC-Zoom (black diamonds). The shape of the density PDF determines how quickly the molecules are produced. The formation timescales for H$_2$ and CO are comparable to (for the LN+PL PDF) or longer than (for the LN PDF) the duration of the simulations which explains why we find that their masses are growing throughout the simulation. The LN PDF consistently underpredicts $M_{\rm H_2}$ with respect to SILCC-Zoom by a factor of $\sim 2$ and produces little CO within the simulation time. This is consistent with the fact that SILCC-Zoom has a much more prominent high-density tail. On the other hand, the LN+PL PDF forms $\rm H_2$ at a rate that is nearly twice as high as SILCC-Zoom and nicely matches the evolution of $M_\mathrm{CO}$ for the first 1.5 Myr, beyond which it saturates at about $50\%$ and $75\%$ of the final value in SILCC-Zoom in the runs with $A_V=0$ and $0.5$, respectively. To give an idea of the long-term evolution of the different species, we also compute their masses at chemical equilibrium for the final snapshot of the simulation (these are indicated with filled ($A_V=0$) and empty ($A_V=0.5$) circles on the right-hand side of the plots). These values never differ substantially from the simulation output at 3 Myr.

Overall, despite the differences in the exact treatment of various physical processes and the chemical network in SILCC-Zoom and our simulations, the predicted masses of the chemical species are in reasonable agreement, in particular when the LN+PL PDF is used. We stress that the goal here is not to achieve a perfect match since, as we mentioned, the sub-grid density
PDF in MC1-HD is quite different from the analytical models used within HYACINTH, which are meant as averages over an ensemble of turbulent molecular clouds in the presence of stellar feedback. Finally, it is worth remembering that, when using HYACINTH as a sub-grid model in cosmological simulations, we switch between the LN and LN+PL PDFs to mimic the combined effects of self-gravity in a turbulent ISM and stellar feedback as explained in Sect.~\ref{sec:stellar_feedback}. The switch takes place on much longer timescales than those investigated in this section and would likely lead to somewhat different results.

%--------------------------------------------
\subsubsection{GML11 simulations}
\label{sec:gml11}
The GML11 simulations track the thermal and chemical evolution of magnetised and supersonically turbulent gas with typical conditions found in molecular clouds. The computational domain is a cube of side length $L = 20 \, \rm pc$ divided into a fixed number of cells. We present results for the runs with a grid-cell size of $0.156 \, \rm pc$  which generate a closely LN density PDF well converged with the exception of the high-density tail. The GML11 simulations consider three different mean densities (100, 300, and 1000 cm$^{-3}$) and use a chemical network (consisting of 218 reactions between 32 species) adapted to model hydrogen, carbon, and oxygen chemistry in molecular clouds. A detailed comparison of the performance in modelling CO chemistry of this more extended network and the simpler one given in NL99 is presented in GC12.

As the boxlength of the GML11 simulations is similar to the cell size we aim to achieve in our future cosmological runs, we model the entire computational volume of GML11 as a single domain in HYACINTH. The DTG is set to $0.01 \, (Z \, / \, \rm Z_{\odot})$ and the elemental abundances of carbon and oxygen are set to $f_{\rm C, tot} = 1.41 \times 10^{-4} (Z/\rm Z_{\odot})$ and $f_{\rm O, tot} = 3.16 \times 10^{-4} (Z/\rm Z_{\odot})$ (same as in GML11). All hydrogen is initially in atomic form while all carbon is in ionised form. The regions are irradiated with an interstellar radiation field (ISRF) of strength $G_0=1.7$ in Habing units and a cosmic ray ionisation rate of $\zeta_{\rm H}=10^{-17}\, \rm s^{-1} \, {\rm H}^{-1}$. We compute the chemical evolution for $5.7 \, \rm Myr$ (the time interval covered by the GML11 runs) and, in Fig.~\ref{fig:mock}, compare the $\rm H_2$ and $\rm CO$ abundances with those from GML11\footnote{The abundances of $\rm C$ and $\rm C^+$ are not presented in GML11.} (as presented in their Figure 5). For $\langle n_{\rm H} \rangle=1000 \, \rm cm^{-3}$, both LN and LN+PL give a reasonable agreement with GML11, while for the lower densities, the $f_{\rm H_2}$ from LN closely matches the $f_{\rm H_2}$ from GML11, but the one from LN+PL is consistently higher. A similar behaviour is observed for $\rm CO$, although LN predicts roughly an order of magnitude lower $f_{\rm CO}$ at $\langle n_{\rm H} \rangle=300 \, \rm cm^{-3}$. In contrast, LN+PL gives a significantly higher $f_{\rm CO}$. At $\langle n_{\rm H} \rangle=100 \, \rm cm^{-3}$, hardly any $\rm CO$ forms in GML11 and in our LN run, while the $f_{\rm CO}$ from LN+PL comprises a small fraction of the total carbon abundance ($\sim 5 \%$). 

As in Fig.~\ref{fig:new_mock_silcc}, on the right hand side of each panel, we show the abundances that would be obtained at chemical equilibrium. For the LN+PL PDF, these are similar to the results obtained after $5.7 \, \rm Myr$, indicating that the molecules form with a characteristic timescale of a few million years. On the contrary, the equilibrium abundances obtained with the LN PDF are significantly higher than those obtained after $5.7 \, \rm Myr$, reflecting a much slower formation rate for the molecules due to the less prominent high-density tail in the PDF.

Overall, HYACINTH with the LN PDF gives H$_2$ abundances in excellent agreement with the GML11 simulations. The situation is more complex for CO, as the formation timescale  of the molecules is very sensitive to the high-density tail of the sub-grid PDF (as we noticed already in Fig.~\ref{fig:new_mock_silcc}) and some fine tuning would be needed to get a good match between the different models. For $\langle n_\mathrm{H}\rangle=100$ and 300 cm$^{-3}$, our results with the LN PDF overestimate the formation time compared to GML11 while those with the LN+PL PDF underestimate it. At $\langle n_\mathrm{H}\rangle=1000$ cm$^{-3}$, CO is always produced at a slightly faster rate in HYACINTH than in the GML11 simulations. This simple test suggests that, at such densities and on such time intervals, HYACINTH overestimates the CO abundance by a factor of $\sim 2$ with respect to GML11. We note, however, that, in a less idealised set-up that accounts for star formation and stellar feedback, the high-density regions would be exposed to much more intense UV radiation than assumed in this test (see, e.g. Fig.~\ref{fig:h2_fraction}) and this would surely affect the resulting CO abundance.

\section{Sample application}
\label{sec:results}
%FIGURES
\begin{figure*}
    \centering    
    \includegraphics[width=0.95\textwidth,trim={3cm 3cm 4cm 0},clip]{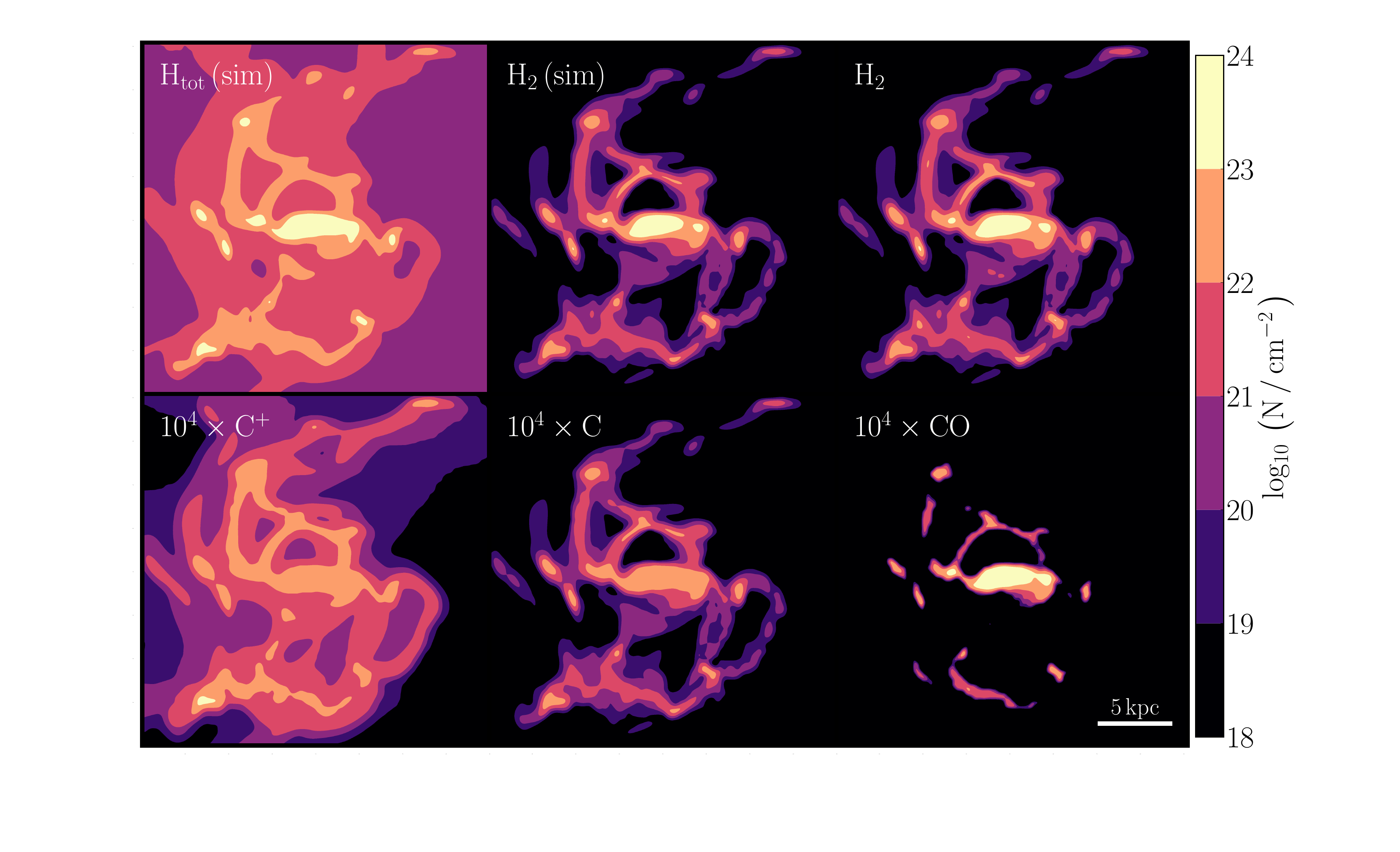}
    \caption{Column density maps of different species in the simulated galaxy after post-processing. The total hydrogen and ${\rm H_2}$ from the simulation are shown in the first two panels. The remaining panels show the column density of ${\rm H_2}$, ${\rm C^+}$ , ${\rm C}$, and ${\rm CO}$ obtained via post-processing. The ${\rm H_2}$ from post-processing (\textit{top row, third panel}) is remarkably similar to the ${\rm H_2}$ from the simulation (\textit{top row, second panel}). ${\rm CO}$ dominates the carbon budget in the highest $N_{\rm H_2}$ regions and is essentially absent when $N_{\rm H_2} \lesssim 10^{21} \, \rm cm^{-2}$. Atomic carbon traces the full extent of the ${\rm H_2}$ distribution, while ${\rm C^+}$ extends farther out to regions devoid of a significant amount of ${\rm H_2}$.}
    \label{fig:ncol6}
\end{figure*}
\begin{figure*}
    \centering
    \includegraphics[width=0.98\textwidth,trim={2cm 0 5cm 0.5cm},clip]
    {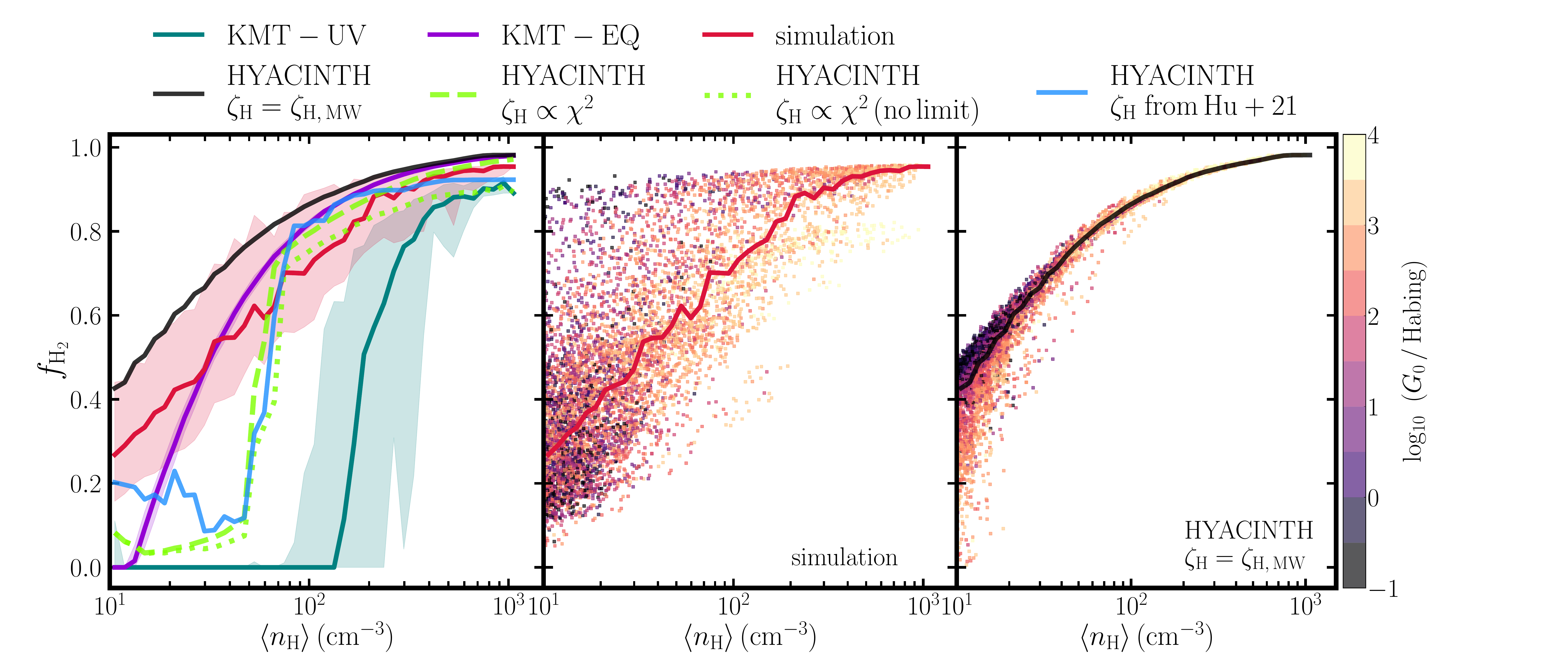}    
    \caption{Comparison of the ${\rm H_2}$ fraction \mbox{($ f_{\rm H_2} \,=\, 2 \langle n_{\rm H_2} \rangle/\langle n_{\rm H} \rangle$)} as a function of $\langle n_{\rm H} \rangle$ using different approaches. \textit{Left:} the equilibrium $f_{\rm H_2}$ from post-processing the T15 galaxy with HYACINTH (black, green, and blue) compared with the $f_{\rm H_2}$ from the simulation (red) and two analytical estimates -- KMT-EQ (blue) and KMT-UV (turquoise). The lines show the median value in a given $\langle n_{\rm H} \rangle$ bin while the shaded area encloses the 16th to 84th percentiles. The solid black line denotes the median $f_{\rm H_2}$ when using a uniform CRIR of $\zeta_{{\rm H}} = 3 \times 10^{-17} \, \rm s^{-1} \, {\rm H}^{-1}$ for HYACINTH. We also show the median $f_{\rm H_2}$ obtained from post-processing when using the CRIR from Sect.~\ref{sec:cr} (i.e. $\zeta_{{\rm H}} \propto \chi^2$) with (dashed green line) and without (dotted green line) the upper limit on the CRIR as well as (solid blue line) the $\zeta_{{\rm H}}-\chi$ relation from \citet[][i.e., $\zeta_{{\rm H}}\propto \chi$, normalised to $\zeta_{{\rm H, \, MW}} = 10^{-16} \, \rm s^{-1} \, H^{-1}$]{hu21}.
    \textit{Middle}: the $f_{\rm H_2}$ from the T15 simulation colour-coded by the strength of the UV field in the LW bands in Habing units, $G_0$ (from the simulation), in each grid cell. The red line is the same as in the left panel. 
    \textit{Right}: the post-processed $f_{\rm H_2}$ (using a uniform $\zeta_{{\rm H}}$) for each grid cell within the simulated galaxy colour-coded by $G_0$. The black line is the same as in the left panel.}
    \label{fig:h2_fraction}
\end{figure*}
\begin{figure*}
    \centering     
    \includegraphics[width=0.98\textwidth,trim={0 0 0 2cm},clip]{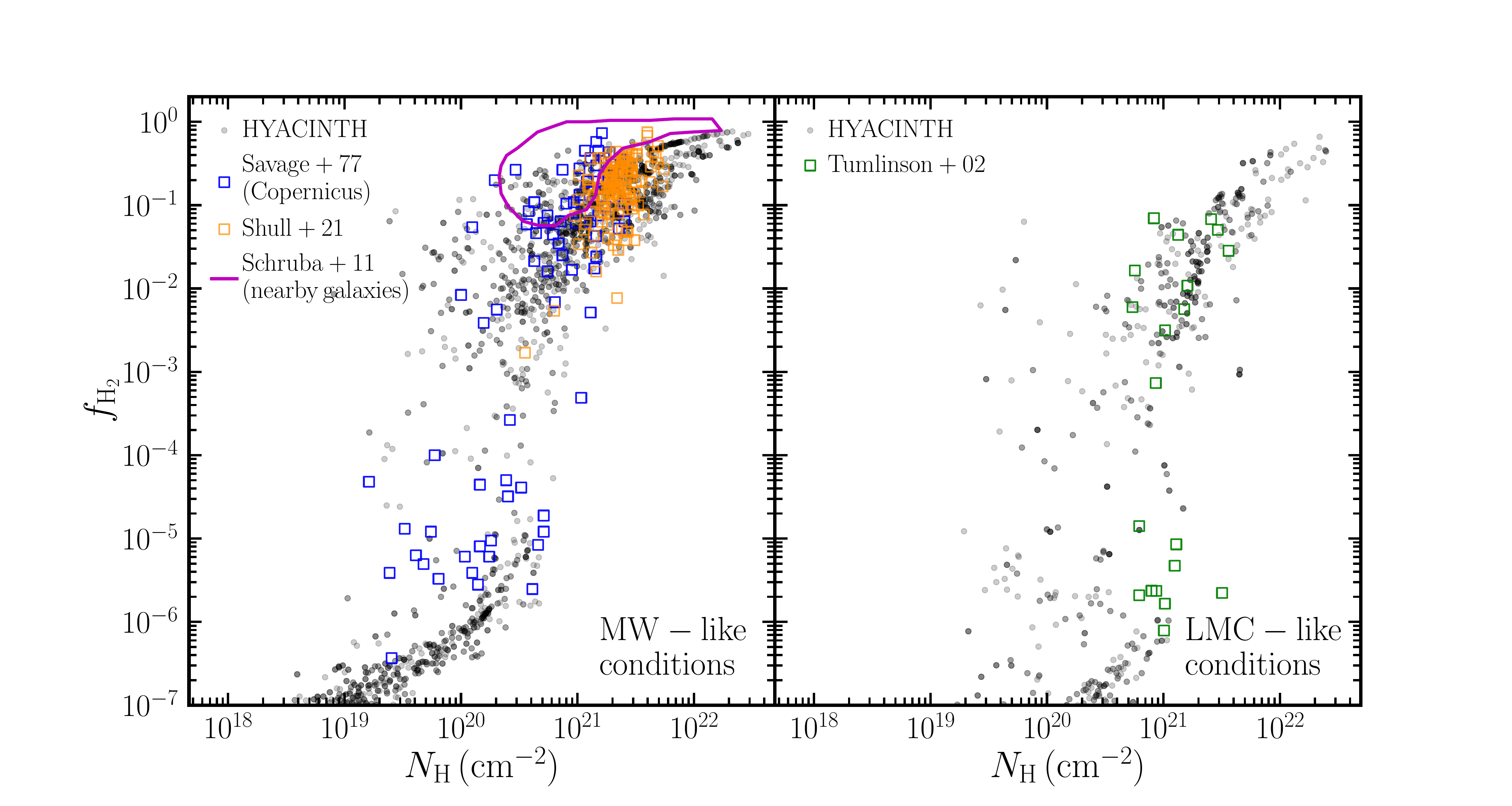}
    \caption{${\rm H_2}$ fraction as a function of the total column density of hydrogen $N_{\rm H}$ = $N_{\rm H \, \textsc{i}} + 2N_{\rm H_2}$, where $f_{{\rm H_2}} = 2N_{\rm H_2} \,  / \, N_{\rm H}$. \textit{Left:} The HYACINTH-post-processed regions with MW-like conditions are shown in black. The observed data are from \cite{savage77} (blue open squares), \cite{shull21} (orange open squares), \cite{schruba11} (magenta dots). For clarity, here we only show a contour enclosing all data points for \cite{schruba11}. 
    \textit{Right:} The HYACINTH-post-processed regions with LMC-like conditions (black) compared to the observed data from \cite{tumlinson02} (green open squares). For HYACINTH data, we use a uniform $\zeta_{\rm H}=\zeta_{\rm H,MW}$ and the column densities are calculated in a position-position-velocity cube of $156 \, \rm pc \, 156 \, \rm pc \times \, 40 \, \rm km \, s^{-1}$, where $156 \, \rm pc $ is the spatial resolution of the AMR grid in the simulation at the redshift of post-processing and $40 \, \rm km \, s^{-1}$ is an upper limit on the observed velocity dispersion in the various absorption measurements used in this comparison. }
    \label{fig:hi_h2_transition}
\end{figure*}
\begin{figure}
    \centering     
    \includegraphics[width=0.48\textwidth,trim={1.5cm 0 0 0},clip]
    {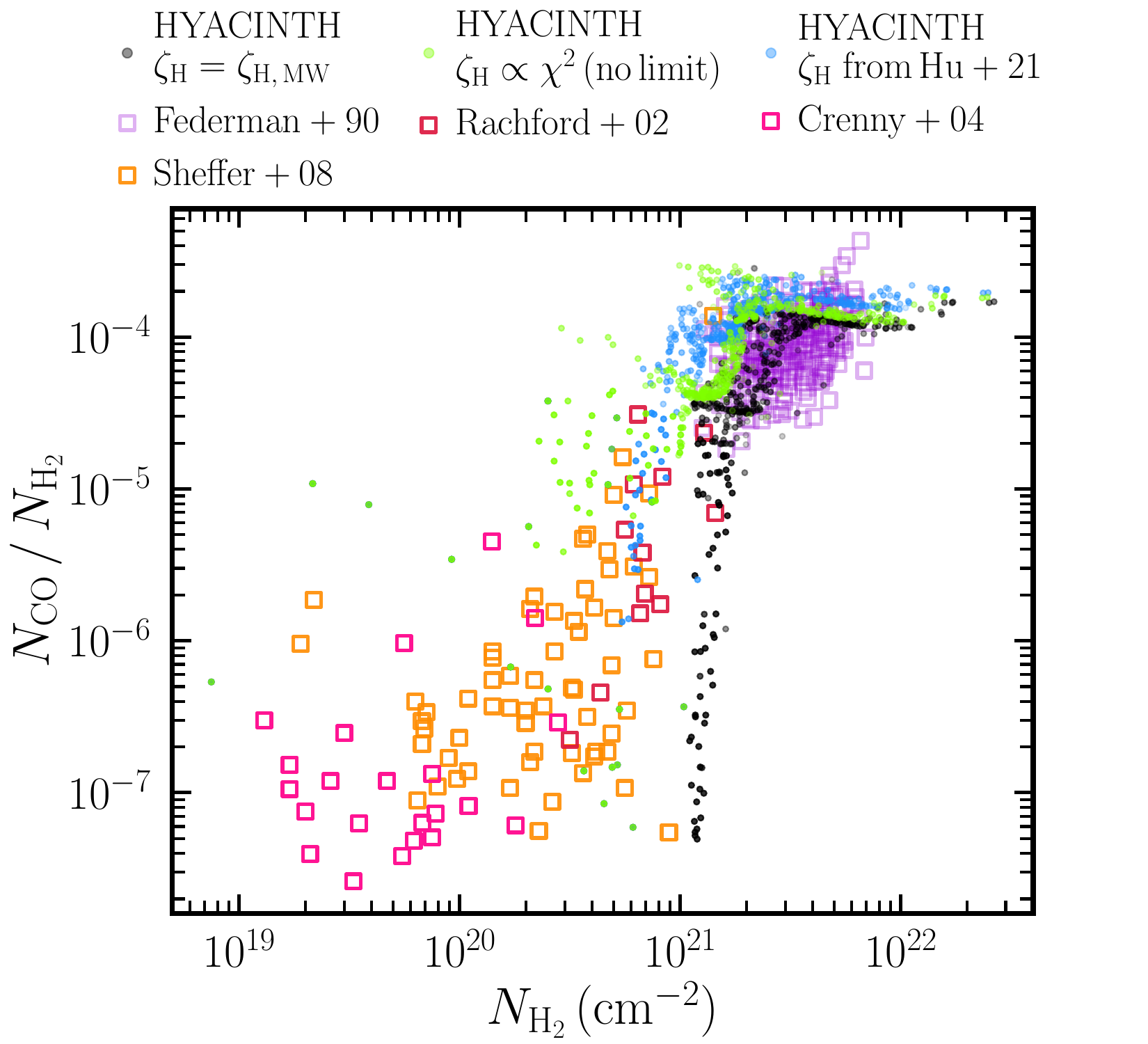}
    \caption{Ratio of the ${\rm CO}$-to-${\rm H_2}$ column density $N_{{\rm CO}}/N_{{\rm H_2}}$ versus $N_{\rm H_2}$ for the post-processed galaxy, compared with observations of (optically) diffuse \citep{rachford02, crenny04, sheffer08} and dark \citep{federman90} molecular clouds in the MW. The black, green, and blue points are from the galaxy post-processed with HYACINTH but using different CRIRs (as indicated in the legend). In each case, the points are selected to have a gas-metallicity $Z \in [0.7,1.3]$ and $\chi \in [0.35,1.3]$ to match the respective values in the observed data. The column densities are calculated in a position-position-velocity cube of $156 \, \rm pc \, 156 \, \rm pc \times \, 5 \, \rm km \, s^{-1}$, to match the velocity dispersion in the absorption measurements. }
    \label{fig:Nco_Nh2}
\end{figure}
In this section, we apply HYACINTH to a simulated galaxy to compare our model predictions with observations of ${\rm H_2}$, ${\rm CO}$, ${\rm C}$, and ${\rm C^+}$ abundances in nearby and high-redshift galaxies. Although our primary goal is to integrate HYACINTH as a sub-grid component within simulations, here we post-process an existing simulation from T15 at $z \sim 2.5$ and compute the equilibrium abundances. The galaxy in T15 was simulated with a modified version of the AMR code \textsc{Ramses} \citep{ramses} including a sub-grid model for ${\rm H_2}$ chemistry and a method to propagate the UV radiation in the Lyman-Werner (LW) bands to nearby cells. ${\rm H_2}$ formation takes place on the surface of dust grains and ${\rm H_2}$ is destroyed by LW photons. In contrast, our chemical network includes additional channels for ${\rm H_2}$ formation and destruction (see Sect.~\ref{sec:h2_chem} and Table~\ref{tab:chem_reactions} for a complete list). Moreover, we adopt a different DTG than T15 (${\rm log_{10}\,(DTG_{T15}) = log_{10}}\,(Z/\rm Z_{\odot}) - 2.0$; see Sect.~\ref{sec:h2_chem} for our DTG). 

An important caveat when applying HYACINTH in post-processing involves the CRIR. In this scenario, the UV field $\chi$ from the simulation is used as an input to compute the equilibrium abundances, meaning that it is assumed that $\chi$ stays constant until equilibrium is reached. Because of the quadratic relationship between the CRIR and $\chi$ in HYACINTH (\ref{sec:cr}), cells in the simulation that have recently undergone star formation will have an extended period of high CRIR. Thus, $\zeta_{{\rm H}} \propto \chi^2$ is not the ideal CRIR for a post-processing application\footnote{It is better suited for dynamically evolving chemistry within simulations where the UV field strength $\chi$ would vary over time.}.Therefore, we adopt a fixed CRIR of $3 \times 10^{-17} \, \rm s^{-1} \, \rm H^{-1}$. Lastly, as applying the model in post-processing does not allow us to switch between the two PDFs, here we only use the log-normal (as in T15). 

Fig.~\ref{fig:ncol6} shows the column-density maps for the species obtained via post-processing as well as the total hydrogen and ${\rm H_2}$ column densities from the simulation. Firstly, we see a remarkable agreement between the $N_{\rm H_2}$ maps from the simulation and post-processing, indicating that the differences in our computation of ${\rm H_2}$ chemistry compared to T15, namely the additional chemical reactions and the equilibrium treatment do not significantly impact the ${\rm H_2}$ column density. This also indicates that the ${\rm H_2}$ in the simulated galaxy is close to equilibrium. From the distribution of the three carbon species, we see that ${\rm CO}$ is the dominant carbon component in the highest $N_{\rm H_2}$ ($\gtrsim 10^{22} \, \rm cm^{-2}$) regions and is hardly present at $N_{\rm H_2} \lesssim 10^{21} \, \rm cm^{-2}$, where $N_{{\rm CO}}$ drops below $10^{15} \, \rm cm^{-2}$. The distribution of atomic carbon (C) shows a great similarity to that of ${\rm H_2}$ both in its extent and the location of peaks. ${\rm C^+}$ is found even more extensively throughout the galaxy, resembling the spread of the total hydrogen ($\rm H_{\rm tot}$). This is because carbon's ionisation energy ($11.6 \, \rm eV$) is slightly lower than that of hydrogen, allowing ${\rm C^+}$ to exist in all ISM phases. We further discuss this in the context of some recent observations in Sect.~\ref{sec:cp_halo}. Another noteworthy feature is that while the density peaks in ${\rm CO}$ and ${\rm C}$ coincide with those in ${\rm H_2}$, the local fluctuations in ${\rm C^+}$ are milder. As ${\rm H_2}$ assists in shielding the former two from UV radiation, their abundances are enhanced in high $N_{\rm H_2}$ regions. On the other hand, ${\rm C^+}$ can exist both in the atomic and molecular phases of the ISM; therefore, we do not see a strong enhancement in $N_{{\rm C^+}}$ with increasing $N_{\rm H_2}$.  

\subsection{Comparison of $\rm H_2$ abundance}
A comparison of the post-processed ${\rm H_2}$ fraction, \mbox{$f_{\rm H_2}=2 \langle n_{\rm H_2} \rangle/\langle n_{\rm H} \rangle$}, with that directly obtained from the simulation is presented in Fig.~\ref{fig:h2_fraction}. We stress that because of the differences in T15 and our chemical network and because we solve for equilibrium, we do not expect to exactly reproduce the dynamically evolved ${\rm H_2}$ abundance from the simulation in every grid cell, but only obtain values similar to those in the simulated galaxy. We also calculate $f_{\rm H_2}$ in each cell using two analytical prescriptions, namely `KMT-EQ' \citep{kmt09} and `KMT-UV' \citep{kmt13}. In the KMT-EQ relation, $f_{\rm H_2}$ is determined by the gas column density and metallicity, and is independent of the strength of the ISRF, $G_0$, by construction. The KMT-UV relation accounts for the effect of UV radiation on $f_{\rm H_2}$ and is sensitive to the ratio $G_0/\langle n_{\rm H} \rangle$. In the left panel, we plot the median $f_{\rm H_2}$ as a function of $\langle n_{\rm H} \rangle$ for the different approaches. Firstly, we see that at $\langle n_{\rm H} \rangle \gtrsim 100 \, \rm cm^{-3}$, the median $f_{\rm H_2}$ from post-processing with a uniform $\zeta_{\rm H}$ (solid black line) agrees very well with the simulation and the KMT-EQ prediction. At lower densities, however, the three show some differences. By $\langle n_{\rm H} \rangle =10 \, \rm cm^{-3}$, the KMT-EQ prediction gradually decreases to $0$, while the simulation has a $f_{\rm H_2} \sim 0.2$. The post-processed $f_{\rm H_2}$ shows a similar trend at low densities but is consistently higher than the simulation. In contrast with all other approaches, KMT-UV predicts a median $f_{\rm H_2}=0$ at $\langle n_{\rm H} \rangle \lesssim 100 \, \rm cm^{-3}$. This median is dominated by the cells that have a very high $G_0/\langle n_{\rm H} \rangle$ value and as such the KMT-UV prescription gives $f_{\rm H_2} = 0$ for all these cells. At $\langle n_{\rm H} \rangle \gtrsim 100 \, \rm cm^{-3}$, the $f_{\rm H_2}$ increases gradually and gives similar results as the other methods. 

Additionally, to demonstrate the impact of using a variable $\zeta_{{\rm H}}$ %\propto \chi^2$ 
in post-processing, we show the $f_{\rm H_2}$ obtained for three different $\zeta_{{\rm H}}-\chi$ relations: (i) $\zeta_{{\rm H}} \propto \chi^2$, but with an upper limit of $3 \times 10^{-14} \, \rm s^{-1} \, H^{-1}$ on $\zeta_{{\rm H}}$ (dashed green curve); (ii) $\zeta_{{\rm H}} \propto \chi^2$ without any upper limit (dotted green curve); (iii) the $\zeta_{{\rm H}}- \chi$ relation from \citet[][solid blue curve]{hu21}. In all three cases, the post-processed $f_{\rm H_2}$ exhibits a sharp drop at $\langle n_{\rm H} \rangle \sim 70 \, \rm cm^{-3}$ and shows a great resemblance to the KMT-UV prediction. We see that removing the upper bound of $3 \times 10^{-14} \, \rm s^{-1} \, H^{-1}$ on the CRIR in our $\zeta_{{\rm H}} \propto \chi^2$ relation (dotted green line) results in a decrease in the median $f_{\rm H_2}$ by $\approx 10 \%$ at $ \langle n_{\rm H} \rangle \geq 100 \, \rm cm^{-3}$. 

In the middle and right panels of Fig.~\ref{fig:h2_fraction}, we show, respectively, the dynamical (i.e. from the simulation) and the equilibrium $f_{\rm H_2}$ (with a uniform $\zeta_{\rm H}$) for each grid cell colour-coded by $G_0$, the strength of the UV field in the LW band in the cell. We see that, at a given $\langle n_{\rm H} \rangle$, the equilibrium $f_{\rm H_2}$ predicted by our model is sensitive to the value of $G_0$ (similar to KMT-UV), with a higher $G_0$ resulting in a lower $f_{\rm H_2}$. In contrast, the simulated $f_{\rm H_2}$ does not show a clear trend with $G_0$, because the UV field varies throughout the formation history of ${\rm H_2}$ in any given region.  Overall, we see that the different approaches for ${\rm H_2}$ chemistry result in varying predictions for $f_{\rm H_2}$. In particular, the abundance computed dynamically in the simulation differs from the equilibrium calculations and shows a larger scatter at all densities. Similar plots when using a variable $\zeta_{\rm H}$  are shown in Fig.~\ref{fig:scatter_h2_fraction} in Appendix~\ref{appendixD}.

\subsection{Comparison with observations}
\subsubsection{The $\mathrm{H} \, \mathrm{\textsc{i}}$ - $\mathrm{H_2}$ transition}
\label{sec:hi_h2}
We compare the ${\rm H}\,\rm \textsc{i}-{\rm H_2}$ transition in the post-processed galaxy (with a uniform $\zeta_{\rm H}$) with the observed one obtained from the absorption spectra of distant quasars and nearby stars in the  MW and Large Magellanic Cloud (LMC). For this, we compute the column densities of ${\rm H_2}$ and total hydrogen within a position-position-velocity data cube of $156\,\rm pc\,\times\, 156\,\rm pc \,\times\, 40 \, \rm km\, s^{-1}$ \footnote{$156\,\rm pc$ is the spatial resolution of the simulation at the redshift of post-processing. The velocity dispersion obtained from the different absorption measurements ranges from $4-35 \, \rm km\,s^{-1}$; here we adopt $40 \, \rm km\, s^{-1}$ as a safe upper limit for these values.}. For a total hydrogen column density  $N_{\rm H} =N_{\rm H \, \textsc{i}}\,+\,2 N_{\rm H_2}$, the ${\rm H_2}$ fraction can be defined as $f_{\rm H_2} = 2\, N_{\rm H_2} / N_{\rm H}$. The results of this comparison are shown in Fig.~\ref{fig:hi_h2_transition}. In the left panel, we show the regions with MW-like conditions, that is $0.6<Z/ \rm Z_{\odot} <1.4$ and $0.6<G_0/1.7<1.4$ and in the right panel, those with LMC-like conditions, that is $0.18<Z/\rm Z_{\odot} <0.42$ and $6<G_0/1.7<14$. The observed data are taken from the Copernicus survey \citep{savage77}, the HERACLES survey of 30 nearby galaxies \citep{schruba11}, and the compilation of several observations of OB stars in the Galactic disc with the Far Ultraviolet Spectrographic Explorer (FUSE) by \cite{shull21}. The LMC data are from \cite{tumlinson02} using a FUSE survey. We see that the ${\rm H}\,\rm \textsc{i}-{\rm H_2}$ transition from post-processing agrees very well with the observed relation, particularly for MW-like conditions (left panel). As evident from both the post-processed and observed data,  the transition is sensitive to the metallicity and the strength of the ISRF: it shifts to higher $N_{\rm H}$ values for LMC-like conditions as compared to MW-like conditions, since the former has $10$ times stronger ISRF than the MW but only about a third of the metals in the MW.

\subsubsection{The relationship between $N_{{\rm H_2}}$ and $N_{{\rm CO}}$}
\label{sec:h2_co}
Fig.~\ref{fig:Nco_Nh2} shows the relation  between the ratio of ${\rm CO}$ to ${\rm H_2}$ column densities ($N_{{\rm CO}}/N_{{\rm H_2}}$) and the ${\rm H_2}$ column density $N_{\rm H_2}$ in the post-processed galaxy for three different CRIRs. For comparison, we include column densities obtained from UV absorption measurements along sight lines towards diffuse and (optically) dark molecular clouds in the MW from \cite{sheffer08}. The $N_{{\rm CO}}/N_{{\rm H_2}}$ values for a compilation of dark-cloud observations by \cite{federman90} are also shown. In these observations, dark clouds are defined as those with visual extinction $A_V \gtrsim 5$, while those with $A_V \lesssim 5$ are defined as diffuse. As these are MW observations, we only plot the post-processed regions with  $0.6<Z/\rm Z_{\odot} <1.4$. Moreover, Figure 7 of \cite{sheffer08} shows that the ISRF values (expressed as $\chi = G_0/1.7$) for these observations range from $\sim 0.5$ to $\sim 10$. Therefore, for a fair comparison, we only select the post-processed regions with $G_0/1.7$ in the range $[0.35,13]$, allowing for a scatter of $30 \, \%$.

At ${\rm H_2}$ column densities above $10^{21} \, \rm cm^{-2}$, our post-processed points (for all CRIRs) span the same region in the $N_{{\rm CO}}/N_{{\rm H_2}}$ versus $N_{{\rm H_2}}$ plane as \cite{federman90}, albeit with significantly less scatter. At lower $N_{{\rm H_2}} \sim 10^{21} \, \rm cm^{-2}$, the $N_{{\rm CO}}/N_{{\rm H_2}}$ ratio shows a sharp drop when using a uniform $\zeta_{{\rm H}} = \zeta_{{\rm H, \, MW }}$, in contrast to the gradual decline seen in the observed data at $N_{{\rm H_2}} \sim 5 \times 10^{20} - 10^{21} \, \rm cm^{-2}$. Conversely, the $N_{{\rm CO}}/N_{{\rm H_2}}$ ratio shows a gradual decline similar to the observed data when using a variable $\zeta_{{\rm H}}$ (CRIR). Nevertheless, both our data and observations exhibit a large scatter in the $N_{{\rm CO}}/N_{{\rm H_2}}$ values at $N_{{\rm H_2}} \in [10^{20}, 10^{21}] \, \rm cm^{-2}$. Moreover our values are in a good agreement with the observed values at these ${\rm H_2}$ column densities. Such large variations in $N_{{\rm CO}}/N_{{\rm H_2}}$ values were also reported previously by \cite{smith14} in their simulation of a MW-like galaxy. At $N_{{\rm H_2}} \lesssim 10^{20}$, our data is sparse and some of our sight lines have a factor $\sim 2$ higher $N_{{\rm CO}}$ compared to those reported by \cite{crenny04} and \cite{sheffer08}. Nevertheless, the majority of our data remains consistent with the observations shown here. Finally, it is worth noting that the regions within our post-processed galaxy with MW-like metallicity predominantly resemble dark clouds while only a few inhabit the region where $N_{{\rm CO}} < 10^{16} \rm cm^{-2}$ and $N_{{\rm H_2}} < 10^{21} \rm cm^{-2}$.

\subsubsection{The abundance of atomic carbon}
\label{sec:h2_c}

The fine structure lines of atomic carbon ${\rm C}\,\rm \textsc{i}$ (corresponding to the $^3P_2 - ^3P_1$ and $^3P_1 - ^3P_0$ transitions) are often used to infer the ${\rm H_2}$ masses of galaxies -- both in the local Universe as well as at high redshifts  \citep[e.g.][]{gerin00, ikeda02, weiss03, weiss05, walter11, valentino18, henriquez-brocal22}. This method requires assuming an atomic carbon abundance relative to ${\rm H_2}$. Several observations have tried to measure this abundance using other independent estimates of the ${\rm H_2}$ mass of a galaxy such as dust emission in the infrared or rotational lines of ${\rm CO}$.  Here we compare the atomic carbon abundance relative to ${\rm H_2}$ for our post-processed galaxy with those found in observations. Our galaxy (post-processed with a uniform $\zeta_{\rm H}$) has $M_{\rm{C}\,\textsc{i}}=6.29\times10^6 \, \rm M_{\odot}$ and $M_{{\rm H_2}}=4.70\times10^{10} \, \rm M_{\odot}$ which implies a galaxy-integrated neutral carbon abundance relative to ${\rm H_2}$, $X_{\rm{C}\,\textsc{i}} = \displaystyle\frac{M_{\rm{C}\,\textsc{i}}}{6\, M_{{\rm H_2}}}$ of $2.23 \times 10^{-5}$ or equivalently $\log_{10} X_{\rm{C}\,\textsc{i}} =-4.65$. The $M_{\rm{C}\,\textsc{i}}$ of our galaxy is similar to that obtained by \cite{matteo_ci} using a different approach. They use a $X_{\rm{C}\,\textsc{i}}$ of $3 \times 10^{-5}$ from \cite{alaghband-zadeh13} and obtained $M_{\rm{C}\,\textsc{i}} = 7.92 \times 10^{6} \, \rm M_{\odot}$ for their simulated galaxy at $z=2$. 

\cite{walter11} carried out a survey of ${\rm C}\,\rm \textsc{i}$ emission in $z>2$ sub-millimetre galaxies (SMGs) and quasar host galaxies using the IRAM Plateau de Bure interferometer and the IRAM 30 m telescope. They obtain a mean $\log_{10} X_{\rm{C}\,\textsc{i}}$ of $-4.08^{+0.16}_{-0.23}$ for their sample of 10 galaxies. \cite{valentino18} carried out a survey of $[X_{\rm{C}\,\textsc{i}}] \, (^3P_1 - ^3P_0)$ in far-infrared-selected main-sequence (MS) galaxies in the COSMOS field at $z \sim 1.2$ with the Atacama Large Millimeter Array (ALMA). They found a mean $\log_{10} X_{\rm{C}\,\textsc{i}}$ of $-4.7\pm0.1$ and $-4.8\pm0.2$, respectively, for $M_{{\rm H_2}}$ estimates based on dust and ${\rm CO}$ measurements for their sample of 12 galaxies. For SMGs at $z\gtrsim2.5$, they obtain a value of $-4.3 \pm 0.2 $  and$-4.2 \pm 0.1$ for $M_{\rm H_2, \rm dust}$ and $M_{\rm H_2, CO}$, respectively. For a sample of 6 MS galaxies at $z = 1-3$ observed as part of the ASPECS, \cite{boogaard20} estimated $X_{\rm{C}\,\textsc{i}} = (1.9 \pm 0.5) \times 10^{-5}$ (i.e. ${\log_{10} X_{\rm{C}\,\textsc{i}} = -4.72^{+0.10}_{-0.13}}$ ). Using a sample of 21 lensed starburst galaxies with $[X_{\rm{C}\,\textsc{i}}]$ detection at $z \sim 1.3-3.5$, \cite{harrington21} found an $X_{\rm{C}\,\textsc{i}}$ of $(6.82 \pm 3.04) \times 10^{-5}$ (i.e. $\log_{10} X_{\rm{C}\,\textsc{i}} = -4.17^{+0.16}_{-0.26}$). A slightly lower value of $\sim -4.5$ is often used in observations of galaxies at $z\gtrsim2$ (e.g. \citealt{weiss03}).  Thus, broadly speaking, the $X_{\rm{C}\,\textsc{i}}$ in our post-processed galaxy is consistent with those obtained for star-forming galaxies at $z\gtrsim 2$. In particular, our $\log_{10} X_{\rm{C}\,\textsc{i}} = -4.65$ is similar to those in MS galaxies, but lower than the values reported for starbursts and SMGs.

\subsubsection{The extended ${\rm C^+}$ distribution}
\label{sec:cp_halo}
\begin{figure}
    \centering     
    \includegraphics[width=0.48\textwidth,trim={0 0 0 0},clip]{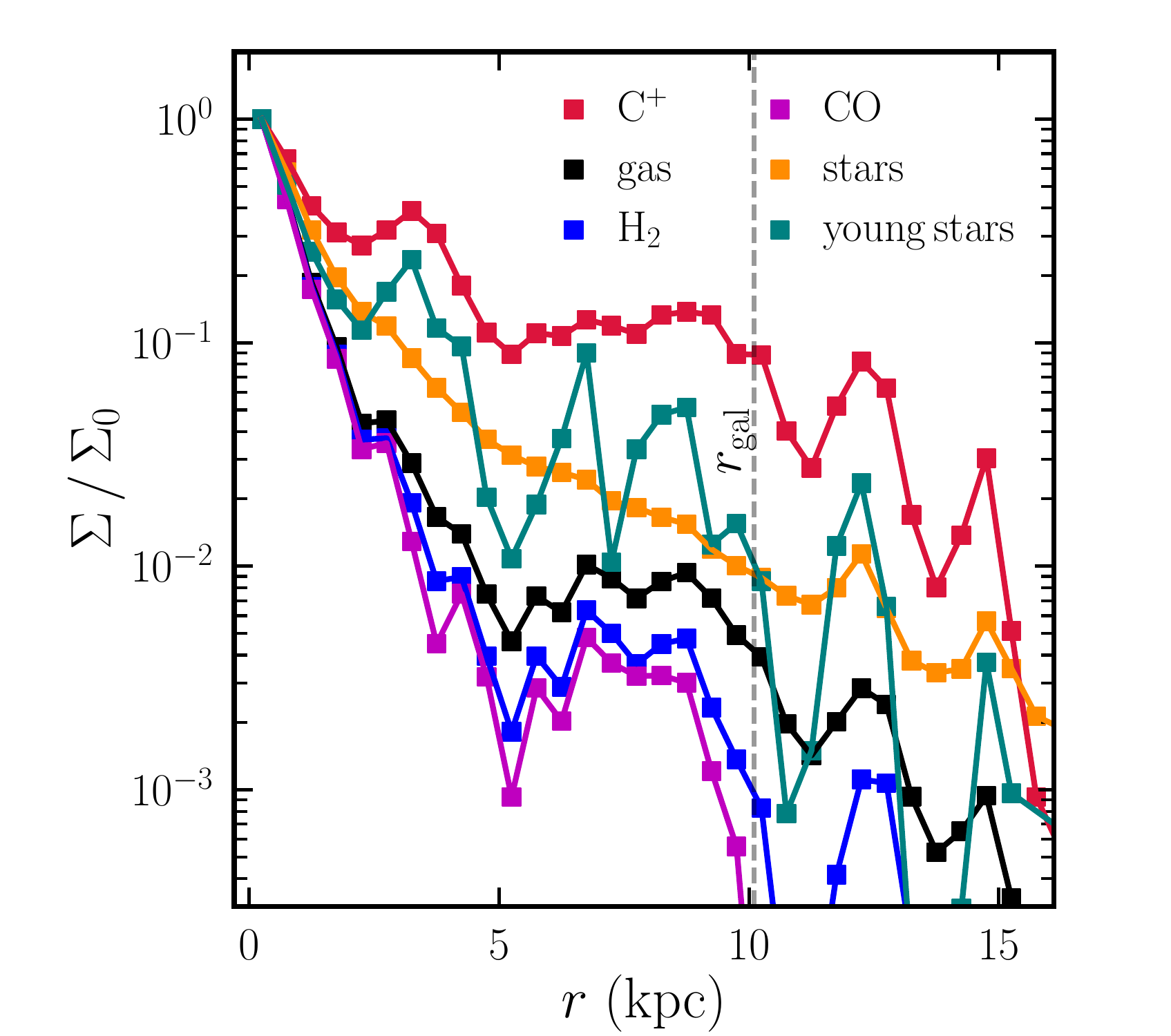}
    \caption{Surface density profiles of gas (black), ${\rm H_2}$ (blue), ${\rm CO}$ (magenta), ${\rm C^+}$ (red), stars (orange), and young stars (teal, with ages less than $10 \, \rm Myr$) within the HYACINTH-post-processed galaxy. For each component, the profiles are normalised by their central value $\Sigma_0$. For each radial bin, we calculate the surface density from the face-on projection from a cylinder with height of $\pm 5 \, \rm kpc$. Among all components, ${\rm C^+}$ and young stars show the slowest decline as we move away from the centre of the galaxy, while ${\rm H_2}$ and ${\rm CO}$ are more concentrated (see also Fig.~\ref{fig:ncol6}). The dashed grey line denotes the size of the galaxy as defined in T15.}
    \label{fig:cp_halo}
\end{figure}
The $[\rm C \, \textsc{ii}]$ fine-structure at 158$\,{\mu} \rm m$ is one of the brightest emission lines in star-forming galaxies and an important coolant of the ISM. The morphology, extent, and kinematics of this line give crucial insights into the different physical processes in galaxies. Several recent $[{\rm C}\, \textsc{ii}]$ observations with ALMA \citep[e.g.][]{nesvadba16, carniani18, fujimoto19, fujimoto20, herrera-camus21, lambert23} have detected the presence of an extended ${\rm C^+}$ reservoir in  galaxies out to $z \sim 7$. These observations find that the effective radius of the $[{\rm C}\, \textsc{ii}]$-emitting  region is $\approx 2-3$ times larger than that of the rest-frame UV-emitting region. Possible explanations for this extended emission include metal-rich outflows, mergers, connection to $\rm Ly \alpha$ halo, and contribution from satellite galaxies, without any clear consensus so far regarding its source \citep{fujimoto19, fujimoto20}. What causes this extended $[{\rm C}\, \textsc{ii}]$ emission remains an open question in the field of galaxy formation.

Fig.~\ref{fig:cp_halo} shows the surface density profiles (normalised to one at the centre) for all gas, ${\rm H_2}$, ${\rm CO}$, ${\rm C^+}$, young stars (age $\leq 10 \, \rm Myr$) and all stars in our post-processed galaxy (with a uniform $\zeta_{\rm H}$) observed face on. The ${\rm H_2}$ and ${\rm CO}$ profiles show a steep decline such that both $\Sigma_{{\rm H_2}}$ and $\Sigma_{{\rm CO}}$ decrease by more than an order of magnitude from the centre out to $2 \, \rm kpc$. The stellar surface density profile shows a similar although less steep decline. In contrast, the ${\rm C^+}$ profile exhibits a gradual decline, maintaining a substantially high surface density at distances extending up to $\sim 10 \, \rm kpc$, beyond which it shows a steep decline.  In this regard, our (normalised) ${\rm C^+}$ profile is similar to the stacked one observed in \cite{fujimoto19} and used to infer the presence of a ${\rm C^+}$ `halo' (which might be an inappropriate name as, in our simulated galaxy, the extended profile arises from the gaseous disc and not from a spheroidal distribution). Young stars also exhibit a relatively slower decline albeit with strong fluctuations. This reflects that UV radiation is essential for ionising carbon. Moreover, this is consistent  with the observed correlation between the $[{\rm C}\, \textsc{ii}]$ line luminosity and the star formation rate \citep{delooze11, delooze14, herrera-camus15, schaerer20}. Also in Fig.~\ref{fig:ncol6}, we saw that the ${\rm C^+}$ distribution strongly mirrors the total hydrogen distribution which is dominated by atomic and ionised hydrogen at larger distances from the centre.  Since we apply HYACINTH in post-processing, all $\rm C^+$ that is farther than the other components is formed in situ and not transported by any dynamical processes such as outflows.

%-----------------------------------------------------------------------------------------
\section{Discussion}
\label{sec:discussion}

It is well known that ${\rm H_2}$ formation can be greatly enhanced by the clumpiness of the ISM on scales typically below the resolution scale of present-day cosmological simulations. Our model accounts for this unresolved density structure by assuming a sub-grid density PDF that varies with the state of star formation in each gas element. This variable PDF is designed to mimic the effect of pre-supernova feedback on the density structure of molecular clouds. In this regard, our approach is alternative to \cite{lupi18} who assume a log-normal PDF with the clumping factor related to the local Mach number. While \cite{lupi18} always assume a log-normal PDF with a continuously varying clumping factor, our approach has a binary nature, where the PDF switches between two distinct shapes - log-normal and log-normal+power-law, each with a constant clumping factor. 

In Sect.~\ref{sec:silcc}, we demonstrated that the chemical abundances from HYACINTH are in a reasonable agreement with high-resolution molecular-cloud simulations. This shows that HYACINTH works reasonably well for modelling molecular hydrogen and carbon chemistry. However, our goal is not to replace extensive chemical networks that would be more accurate for modelling chemistry in ISM-scale simulations, but rather the model is tailored to follow hydrogen and carbon chemistry in large cosmological simulations that have a resolution of $20-200 \, \rm pc$. 

In Sect.~\ref{sec:results}, we have shown a post-processing application for a galaxy simulation (at $z \sim 2.5$) with a spatial resolution of $156 \, \rm pc$. The resulting equilibrium ${\rm H_2}$ abundance are in agreement with the commonly adopted KMT-EQ relation and the ${\rm H_2}$ column density map shows a great similarity to the one obtained directly from the simulation (accounting for the non-equilibrium ${\rm H_2}$ abundance). This shows that our approach gives reasonable results on larger scales ($\sim 100-200 \, \rm pc$) that barely resolve individual molecular clouds.

An important caveat about using HYACINTH is that the formation of ${\rm H_2}$ in HYACINTH relies on the presence of dust, similar to most other approaches in the literature (e.g. KMT). Therefore, it cannot be applied to gas with arbitrarily low metallicities ($Z \lesssim 10^{-3} \, \rm Z_{\odot}$). As a result it cannot be applied to pristine gas such as in simulations of population III stars and early galaxy formation ($z \gtrsim 15$,  see e.g. \citealt{hirano15, lenoble24}). Accurately modelling hydrogen chemistry in such (proto) galaxies requires accounting for three-body reactions and more importantly a high spatial resolution to resolve the densities where these reactions are efficient ($n_{\rm H} \gtrsim 10^8 \, \rm cm^{-3}$). One workaround for cosmological simulations that do not resolve these early objects is to impose a metallicity floor that mimics the metal enrichment by the first stars \citep[i.e. the Population III stars; see e.g.][]{kuhlen12, kuhlen13, matteo_h2, serra22}.  

%-----------------------------------------------------------------------------------------
\section{Summary}
\label{sec:summary}
We present a new sub-grid model, HYACINTH, that can be embedded into cosmological simulations for calculating the non-equilibrium abundances of ${\rm H_2}$ and its carbon-based tracers, namely ${\rm CO}$, ${\rm C}$, and ${\rm C^+}$. HYACINTH comprises a variable sub-grid density PDF to capture the unresolved density structure in simulations and a simplified chemical network for hydrogen and carbon chemistry. These simplifications were introduced to make the network highly efficient for use in large-scale simulations. Additionally, a metallicity-dependent temperature-density relation, based on high-resolution simulations of the star-forming ISM, was used to assign a temperature to each sub-grid density. 

We compared HYACINTH against more sophisticated approaches in the literature for modelling hydrogen and carbon chemistry including two extensive chemical networks (NL99 and G17) and a PDR code, using a one-dimensional semi-infinite slab setup (Fig.~\ref{fig:nl99_g17}). HYACINTH reproduced the $f_{\rm H_2}$-$A_V$ relation from these methods highlighting that $\rm H_2$ chemistry is insensitive to the exact treatment of carbon chemistry. Moreover, despite its simplicity and size, HYACINTH captured the ${\rm C^+} \rightarrow {\rm C} $ and ${\rm C} \rightarrow {\rm CO}$ transitions very well, as predicted by more complex approaches (see Sect.~\ref{sec:nl99_g17} for a detailed discussion).

We compared the predictions from HYACINTH with high-resolution molecular-cloud simulations -- the SILCC-Zoom simulations at $Z= \rm Z_{\odot}$ and the GML11 simulations at $Z=0.1 \, \rm Z_{\odot}$. We find reasonable agreement between the chemical abundances from HYACINTH and those from the simulations. We further find that our log-normal+power-law PDF shows better agreement with the SILCC-Zoom simulations while the log-normal PDF performs better for the GML11 runs, in alignment with the different density structures shown by the two simulations.

Finally, we applied HYACINTH to a simulated galaxy at $z \sim 2.5 $ from T15 in post-processing and compared it directly with observations. For regions in the post-processed galaxy with MW-like conditions, the ${\rm H}\,\rm \textsc{i}$-to-${\rm H_2}$ transition in the $f_{\rm H_2}$-$N_{\rm H}$ plane aligns very well with that from observations. The observed $f_{\rm H_2}$ and $N_{\rm H}$ were obtained from measurements of the absorption spectra of quasars and nearby stars in sightlines towards MW molecular clouds. The same is also true for LMC-like regions (Fig.~\ref{fig:hi_h2_transition}). Additionally, the values of $N_{{\rm CO}} / N_{\rm H_2}$ versus $N_{\rm H_2}$ in MW-like regions are consistent with observations of MW molecular clouds (Fig.~\ref{fig:Nco_Nh2}). It is worth noting that most of our post-processed regions resemble the observed (optically) dark molecular clouds in the $N_{{\rm CO}}/N_{\rm H_2}-N_{\rm H_2}$ plane. Furthermore, the relative abundance of atomic carbon to molecular hydrogen ($X_{\rm{C}\,\textsc{i}}$) in our post-processed galaxy consistently matches the $X_{\rm{C}\,\textsc{i}}$ values found in star-forming galaxies at redshifts $z\gtrsim 1$. Based on surface density profiles of the different baryonic components within the post-processed galaxy, we find an excess of ${\rm C^+}$ at large distances from the galaxy centre, similar to those found in observations \citep[e.g.][]{fujimoto19}.

In a forthcoming paper (paper II; Khatri et al., in prep.), we will present a suite of cosmological simulations using HYACINTH to model the non-equilibrium abundances of ${\rm H_2}$ and its tracers in high-redshift  ($z\gtrsim 2$) galaxies. This will open up the possibility of addressing fundamental questions such as what the contribution is of low-mass galaxies to the global ${\rm H_2}$ budget at high $z$, what regulates the molecular gas fraction in galaxies, and for what physical conditions, environments, and galaxies are ${\rm CO}$, ${\rm C}$, and ${\rm C^+}$ all reliable tracers of molecular gas. 

%--------------------------------------------------------------------

\begin{acknowledgements}
The authors thank an anonymous reviewer and Pascal Oesch for suggestions and comments that helped improving the paper. They gratefully acknowledge the Collaborative Research Center 1601 (SFB 1601 sub-projects C5, B1, B4) funded by the Deutsche Forschungsgemeinschaft (DFG, German Research Foundation) – 500700252. Early stages of this work were carried out within the Collaborative Research Centre 956 (SFB 956 sub-projects C4, C5, C6), funded by the DFG – 184018867. PK is a part of the International Max Planck Research School in Astronomy and Astrophysics, the Bonn Cologne Graduate School, and a guest at the Max Planck Institute for Radio Astronomy in Bonn. DS acknowledges funding from the programme ``Profilbildung 2020'', an initiative of the Ministry of Culture and Science of the State of North Rhine-Westphalia. CP is grateful to SISSA, the University of Trieste, and IFPU, where part of this work was carried out, for hospitality and support. 
\end{acknowledgements}


\bibliographystyle{aa}
\bibliography{example} % if your bibtex file is called example.bib

\begin{appendix}
\section{Reactions in the chemical network}
\label{appendixA}
%\begin{multicols}{1}
%\begin{strip}
%\FloatBarrier
Table~\ref{tab:chem_reactions} gives a list of all chemical reactions in HYACINTH. For each of these, we use updated reaction rates from UMIST \citep{umist12} and KIDA \citep{wakelam12} databases. 
\begin{table*}[b]
    \caption{Reactions in our chemical network along with the rate coefficients.}
    \centering
    %\resizebox{\textwidth}{!}{%
    \begin{tabular}
    {ccccc}
         \hline\hline
         & \textbf{Reaction} & \textbf{Type} & \textbf{Rate coefficient} & \textbf{Reference}  \\
         \hline
         1 & ${\rm C^+} + {\rm H_2}\rightarrow {\rm CH_{\rm x}} + {\rm H}$ & ${\rm H_2}$ destruction, ${\rm CH_{\rm x}}$ formation & 
         $2.31 \times 10^{-13} \, T^{-1.3} \, \exp{ (-23/T)}$ 
         & 1,19
         \\
         2 & ${\rm C^+} + {\rm H_2}\rightarrow {\rm C} + 2 {\rm H}$ & ${\rm H_2}$ destruction, ${\rm C}$ formation & 
         $0.99 \times 10^{-13} \, T^{-1.3} \, \exp{ (-23/T)}$
         & 1 
         \\
         3 & ${\rm H_3^+} + {\rm C} \rightarrow {\rm CH_{\rm x}} + {\rm H_2}$ &  ${\rm CH_{\rm x}}$ formation &
         $1.04 \times 10^{-9} \, (300/T)^{0.00231} + $ &
         \\
         &&& $T^{-1.5} \, \Sigma_{i=1}^{4} \, c_i \, \exp{(-T_i/T)}$; &
         \\
         &&& $c_i = [3.4 \times 10^{-8}, 6.97 \times 10^{-9},$ &\\
         &&& $1.31 \times 10^{-7}, 1.51 \times 10^{-4}]$,&
         \\
         &&& $T_i = [7.62, 1.38, 26.6,8110] $
         &2, 3  
         \\
         4 & ${\rm H_3^+} + {\rm O} \rightarrow {\rm OH_{\rm x}} + {\rm H_2}$ & ${\rm OH_{\rm x}}$ formation &
        $1.99 \times 10^{9} \, T^{-0.190} $  
         & 3, 4
         \\
         5 & ${\rm CH_{\rm x}} + {\rm O} \rightarrow {\rm CO} + {\rm H} $ & ${\rm CO}$ formation &
         $7.7\times 10^{-11}$  
         & 1
         \\
         6 & ${\rm OH_{\rm x}} + {\rm C} \rightarrow {\rm CO} + {\rm H}$ & ${\rm CO}$ formation &
         $7.95 \times 10^{-10} \, T^{-0.339} \, \exp{ (0.108/T)}$  
         &1, 5
         \\
         7 & ${\rm H_3^+} + {\rm e^-}\rightarrow {\rm H_2} + {\rm H} $ & Dissociative recombination & 
         $4.54 \times 10^{-7} \times T^{-0.52}$
         &11, 12
         \\
         8 & ${\rm H_3^+} + {\rm e^-}\rightarrow \rm 3{\rm H} $ & Dissociative recombination &
         $8.46 \times 10^{-7} \times T^{-0.52}$
         &11, 12
         \\
         9 & ${\rm C^+} + {\rm e^-}\rightarrow \rm C + \gamma $ & Radiative recombination &
         $\frac{2.995 \times 10^{-9}}{\alpha (1+\alpha)^{1-\gamma} \, (1+\beta)^{1+\gamma}}$ & 13,14
         \\
         %&&& $\alpha = \sqrt{\frac{T}{6.67 \times 10^{-3}}}$, 
         %$\beta = \sqrt{\frac{T}{1.9436 \times 10^{6}}}$, &
         %\\
         %&&& $\gamma = 0.7849 + 0.1597 \exp{(49550/T)}$
         %&13, 14
         %\\
         %\hline
         10 & ${\rm C}+ \gamma\rightarrow {\rm C^+} + \rm {\rm e^-} $ & Photoionisation & 
         $3.5 \times 10^{-10} \, (G_0/1.7) \, $ & \\
         &&& $\exp{(-3.76 \, A_V)} \, f_{s,{\rm C}}(N_{{\rm C}}, N_{{\rm H_2}})$
         &6, 7  
         \\
         11 & ${\rm H_2} + \gamma\rightarrow 2 {\rm H}$ & Photodissociation &
         $4.2 \times 10^{-11} \, G_0 \, $ \\
         &&& $\exp{(-4.18 \, A_V)} \, f_{s,{\rm H_2}}(N_{{\rm H_2}})$
         &6, 7  
         \\
         12 & ${\rm CO} + \gamma\rightarrow \rm C + O $ & Photodissociation &
         $2.4 \times 10^{-10} \, (G_0/1.7) \,$ \\
         &&& $\exp{(-3.88 \, A_V)} \, f_{s,{\rm CO}}(N_{{\rm CO}}, N_{{\rm H_2}})$
         &6, 8  
         \\
         13 & ${\rm CH_{\rm x}} + \gamma\rightarrow {\rm C} + {\rm H} $ & Photodissociation & 
         $9.1 \times 10^{-10} \, (G_0/1.7) \, \exp{(-2.12 \, A_V)}$
         &6 
         \\
         14 & ${\rm OH_{\rm x}} + \gamma\rightarrow \rm O + H $ & Photodissociation &
         $3.8 \times 10^{-10} \, (G_0/1.7) \, \exp{(-2.66 \, A_V)}$
         &6 
         \\
         %\hline
         15 & ${\rm H_2} + \rm CR \rightarrow {\rm H_2}^+ + {\rm e^-} $ & Cosmic-ray ionisation & 
         $2 \zeta_{\rm H}$
         & 20
         \\
         16 & ${\rm C} + \rm CR \rightarrow {\rm C^+} + {\rm e^-}$ & Cosmic-ray ionisation &
         $3.85 \zeta_{{\rm H}}$
         & 20
         \\
         17 & ${\rm CO} + \rm CR \rightarrow \rm {\rm CO^+} + O $& Cosmic-ray ionisation &
         $6.52 \zeta_{\rm H}$
         & 20,22
         \\         
         %\hline
         18 & ${\rm C^+} + {\rm e^-} + \rm grain \rightarrow {\rm C} $ & Grain-assisted recombination &
         $4.558 \times 10^{-13} \, [1+6.089 \times 10^{-3} \, \psi^{1.128} \times$&
         \\
         &&& $(1+433.1\, T^{0.04845} \, \psi^{-0.8120 - 1.333 \times 10^{-4} \, {\rm ln} T})]^{-1}; $ & 15
         %\\
         %&&& $\psi = \frac{G_0 \, \exp{(-1.87 \, A_V) \, \sqrt{T}}}{n_{{\rm e^-}} / \rm cm^{-3}}$
         \\
         19 & ${\rm H} + {\rm H} + \rm grain\rightarrow {\rm H_2} + \rm grain $ & Grain-assisted formation of ${\rm H_2}$ &
         $3 \times 10^{-17}$
         &17, 18
         \\
         %\hline
         20 & ${\rm He} + \rm CR \rightarrow {\rm He^+} + {\rm e^-}$
         & Cosmic-ray ionisation
         &$1.1 \zeta_{\rm H}$
         & 20,22
         \\
         21 & ${\rm He^+} + {\rm H_2}\rightarrow {\rm H^+} + {\rm He} + {\rm H}$
         & Dissociative charge exchange
         & $ 1.26 \times 10^{-13} \, \exp{(-22.5/T)}$
         &3,27
         \\
         22 & ${\rm He^+} + {\rm H_2}\rightarrow {\rm He} + {\rm H_2^+}$
         & Charge exchange
         &$7.2 \times 10^{-15}$
         & 25, 26
         \\
         23 & ${\rm He^+} + {\rm CO}\rightarrow {\rm He} + {\rm C^+} + {\rm O}$
         & $\rm CO$ destruction
         &$1.6 \times 10^{-9}$
         & 23
         \\
         24 & ${\rm He^+} + {\rm e^-}\rightarrow {\rm He} + \gamma$
         & Radiative recombination
         &$10^{-11} \, T^{-0.5} \times [11.19 - 1.676 \, \rm log_{10} T $ 
         & \\ 
          &&&$ - 0.2852 \, (\rm log_{10} T)^2 + 0.04433 \, (\rm log_{10} T)^3]$ & 21,22
         \\
         25 & ${\rm He^+} + {\rm e^-} + \rm grain \rightarrow {\rm He} + \gamma$
         & Grain-assisted recombination
         & $5.572 \times 10^{-14} \, [1 +3.185 \times 10^{-7} \psi^{1.512} \times$
         & \\
         &&& $(1+5115 \, T^{3.903 \times 10^{-7}} \psi^{-0.4956-5.494 \times 10^{-7} \ln{T}})]^{-1} \, ;$ & 
         \\
         %&&& $\psi$ is the same as in reaction 18 & 15 
         %\\
         26 & ${\rm H_3^+} + {\rm CO}\rightarrow {\rm HCO^+} + {\rm H_2} $
         & $\rm CO$ destruction, $\rm HCO^+$ formation
         &$1.7 \times 10^{-9}$
         &1
         \\
         27 & ${\rm C^+} + {\rm OH_{\rm x}}\rightarrow {\rm HCO^+} $
         & $\rm C^+$ destruction, $\rm HCO^+$ formation
         &$9.15 \times 10^{-10} \, (0.62+45.41 \, T^{-1/2}$
         &1
         \\
         28 & ${\rm HCO^+} + {\rm e^-}\rightarrow {\rm CO} + {\rm H}$
         & Dissociative recombination
         &$1.06 \times 10^{-5} \, T^{-0.64}$
         &24
         \\
         29 & ${\rm HCO^+} + \gamma \rightarrow {\rm CO} + {\rm H^+}$
         & Photodissociation 
         & $5.4 \times 10^{-12} \, (G_0/1.7) \, \exp{(-3.3\, A_V)}$
         & 26
         \\
         \hline
    \end{tabular}
    \tablefoot{The rate coefficients are in $\rm cm^3 \, s^{-1}$ for reactions 1-9, 21-24, and 26-28; in $\rm s^{-1}$ for reactions 10-17, 20, and 29; in $\rm cm^3 \, s^{-1} \, Z_{d}^{-1}$ for 18, 19, and 25, where $Z_{\rm d}$ is the dust abundance relative to the solar neighbourhood value of $0.01$. $\zeta_{{\rm H}}$ is the value of the cosmic ray ionisation rate in units of $\rm s^{-1} \, {\rm H}^{-1}$; $G_0$ is the flux of the radiation field in the Habing units; $A_V$ is the visual extinction defined in Eq. (\ref{eq:Av}). In reaction 9, $\alpha = \sqrt{T/(6.67 \times 10^{-3})}$, $\beta = \sqrt{T/(1.9436 \times 10^{6})}$, and $\gamma = 0.7849 + 0.1597 \exp{(49550/T)}$. In reactions 18 and 25, $\psi = G_0 \, \exp{(-1.87 \, A_V) \, \sqrt{T}}/(n_{{\rm e^-}} / \rm cm^{-3})$. We note that although \cite{heays17} provided an updated value of the unshielded photodissociation rate for reaction 11 of $5.7 \times 10^{-11} \, (G_0/1.7)$, here we use the old rate from \cite{draine_bertoldi}, that is $\approx 25 \%$ lower, for a fair comparison with T15. Reactions 20-29 are only part of extended HYACINTH. }
    \tablebib{(1) \cite{wakelam12};
    (2) \cite{vissapragada16};
    (3) \cite{GOW17};
    (4) \cite{ruette16};
    (5) \cite{zanchet09};
    (6) \cite{heays17};
    (7) \cite{draine_bertoldi};
    (8) \cite{visser09};
    (9) \cite{glassgold74};
    (10) \cite{liszt03};
    (11) \cite{mccall04}; 
    (12) \cite{woodall07};
    (13) \cite{badnell03};
    (14) \cite{badnell06};
    (15) \cite{weingartner01};
    (16) \cite{draine03};
    (17) \cite{wolfire08};
    (18) \cite{hollenbach2012};
    (19) \cite{anicich86};
    (20) \cite{umist99};
    (21) \cite{hummer98};
    (22) \cite{glover10};
    (23) \cite{kim75};
    (24) \cite{geppert05};
    (25) \cite{barlow84};
    (26) \cite{umist12};
    (27) \cite{schauer89}
    .}
    \label{tab:chem_reactions}
\end{table*}

\clearpage 
The fractional abundance $f_X$ of a species $X$  relative to hydrogen is defined as $f_X=n_X/n_{\rm H}$, except for $\rm H_2$ and $\rm H_3^+$, where $f_{\rm H_2} = 2 n_{\rm H_2} / n_{\rm H}$ and $f_{\rm H_3^+} = 3 n_{\rm H_3^+} / n_{\rm H}$. For ${\rm H_2}$, ${\rm CO}$, and ${\rm C^+}$, the differential equations for each sub-grid density $n_{\rm H}$ can be written as\footnote{Atomic carbon is denoted as ${\rm C}\, \textsc{i}$ here.}
\begin{multline}
    \frac{\mathrm{d} f_{{\rm H_2}}}{\mathrm{d} t}\,=\,
    2k_{{\rm H_2},{\rm gr}}\,f_{{\rm H}\,\rm \textsc{i}}\,n_{\rm H}\,-\,
    k_{\gamma,{\rm H_2}}\,f_{{\rm H_2}}\,-\,
    k_{{\rm c.r.,{\rm H_2}}}\,f_{{\rm H_2}} \\ \,-\,
    n_{\rm H}\,k_{{\rm C^+},{\rm H_2}}\,f_{{\rm C^+}}\,f_{{\rm H_2}}\,\alpha_{{\rm CH_{\rm x}}}\,+\,
    \frac{2}{3}\,n_{\rm H}\,k_{{\rm e^-},{\rm H_3^+}}\,f_{{\rm e^-}}\,f_{{\rm H_3^+}} \, ;
    \label{eq:de_h2}
\end{multline}
\begin{multline}
    \frac{\mathrm{d} f_{{\rm CO}}}{\mathrm{d} t}\,=\,
    \frac{1}{3}\,n_{\rm H}\,k_{{\rm O}\,\rm \textsc{i},{\rm H_3^+}}\,f_{{\rm O}\,\rm \textsc{i}}\,f_{{\rm H_3^+}}\,\alpha_{{\rm OH_{\rm x}}}\,+\,
    \frac{1}{3}\,n_{\rm H}\,k_{{\rm C}\, \textsc{i},{\rm H_3^+}}\,f_{\rm{C}\,\textsc{i}}\,f_{{\rm H_3^+}}\,\alpha_{{\rm CH_{\rm x}}} \\ \,+\,
    \frac{1}{2}\,n_{\rm H}\,k_{{\rm C^+},{\rm H_2}}\,f_{{\rm C^+}}\,f_{{\rm H_2}}\,\alpha_{{\rm CH_{\rm x}}}\,-\,
    k_{\gamma,{\rm CO}}\,f_{{\rm CO}} \, ;
    \label{eq:de_co}
\end{multline}
\begin{multline}
    \frac{\mathrm{d} f_{{\rm C^+}}}{\mathrm{d} t}\,=\,
    \,-\,n_{\rm H}\,k_{{\rm C^+},{\rm e^-}}\,f_{{\rm C^+}}\,f_{{\rm e^-}}
    \,-\,n_{\rm H}\,k_{{\rm C^+},{\rm e^-},{\rm gr}}\,f_{{\rm C^+}}\,f_{{\rm e^-}}\\ \,-\,
    \frac{1}{2}\,n_{\rm H}\,k_{{\rm C^+},{\rm H_2}}\,f_{{\rm C^+}}\,f_{{\rm H_2}}\,\alpha_{{\rm CH_{\rm x}}}\,+\,
    k_{\gamma,\rm{C}\,\textsc{i}}\,f_{\rm{C}\,\textsc{i}} \, .
    \label{eq:de_cp}
\end{multline}
The reaction rate coefficient $k$ for each reaction is listed in Table~\ref{tab:chem_reactions}. In the above equations, $\alpha_{{\rm CH_{\rm x}}}$ ($\alpha_{{\rm OH_{\rm x}}}$) is the branching ratio (see Sect.~\ref{sec:branching_ratios}) for the formation of ${\rm CH_{\rm x}}$ (${\rm OH_{\rm x}}$) when $\rm H_3^+$ reacts with atomic carbon (oxygen). 
The species $\rm O_2$, $\rm H_2O$, $\rm OH$, $\rm OH ^+$, $\rm H_2 O^+$, and $\rm H_3$O$^+$  are collectively referred to as ${\rm OH_{\rm x}}$ in the chemical network (as originally done in NL99 based on the argument that these species have similar chemical reactions and reaction rates). Similarly, $\rm CH$, $\rm CH_2$, $\rm CH ^+$, and $\rm CH_2^+$ are collectively referred to as ${\rm CH_{\rm x}}$. For ${\rm H_3^+}$, we assume a local equilibrium (i.e. at each sub-grid density) between the formation and destruction pathways such that:
\begin{equation}\label{eq:h3p1}
\begin{aligned}
        \frac{{\rm d} f_{{\rm H_3^+}}}{{\rm d} t} \,=\, 
    &\frac{3}{2}\,k_{{\rm c.r.,{\rm H_2}}}\,f_{{\rm H_2}}\,
    -n_{\rm H} \,k_{\rm{C}\,\textsc{i},{\rm H_3^+}}\,f_{\rm{C}\,\textsc{i}}\,f_{{\rm H_3^+}}
    \\
    &-n_{\rm H} \,k_{{\rm O}\,\rm \textsc{i},{\rm H_3^+}}\,f_{{\rm O}\,\rm \textsc{i}}\,f_{{\rm H_3^+}}
    -n_{\rm H} \,k_{{\rm e^-},{\rm H_3^+}}\,f_{{\rm e^-}}\,f_{{\rm H_3^+}}
    \,=\, 0 \, ,
\end{aligned}
\end{equation}
which implies
\begin{equation}\label{eq:h3p2}
    f_{{\rm H_3^+}}\,=\,\frac{3}{2\,n_{\rm H}}\,
    \frac{k_{{\rm c.r.,{\rm H_2}}}\,f_{{\rm H_2}}}
    {k_{\rm{C}\,\textsc{i},{\rm H_3^+}}\,f_{\rm{C}\,\textsc{i}}\,+k_{{\rm O}\,\rm \textsc{i},{\rm H_3^+}}\,f_{{\rm O}\,\rm \textsc{i}}\,+k_{{\rm e^-},{\rm H_3^+}}\,f_{{\rm e^-}}\,} \, .
\end{equation}
Similarly, an equilibrium abundance of ${\rm OH_{\rm x}}$ and ${\rm CH_{\rm x}}$ implies
\begin{equation}\label{eq:ohx1}
\begin{aligned}
    \frac{{\rm d} f_{{\rm OH_{\rm x}}}}{{\rm d} t} \,=\, 
    & \frac{1}{3} k_{{\rm O}\,\rm \textsc{i},{\rm H_3^+}}\,f_{{\rm O}\,\rm \textsc{i}} f_{{\rm H_3^+}}  n_{\rm H} \, \\
    & - k_{{\rm C}\,\rm \textsc{i}, {\rm OH_{\rm x}}} f_{\rm{C}\,\textsc{i}} f_{{\rm OH_{\rm x}}} n_{\rm H} \,
    - k_{\gamma, {\rm OH_{\rm x}}} \, f_{{\rm OH_{\rm x}}}
    \,=\, 0 \, ;
\end{aligned}
\end{equation}
\begin{equation}\label{eq:chx1}
\begin{aligned}
    \frac{{\rm d} f_{{\rm CH_{\rm x}}}}{{\rm d} t} \,=\, 
    \frac{1}{3} k_{{\rm C}\,\rm \textsc{i}, {\rm H_3^+}} f_{\rm{C}\,\textsc{i}} f_{{\rm H_3^+}} n_{\rm H} \,
    + \frac{1}{2} k_{{\rm C^+}, {\rm H_2}} f_{{\rm C^+}} f_{{\rm H_2}} n_{\rm H} \, \\
    - k_{{\rm O}\,\rm \textsc{i}, {\rm CH_{\rm x}}} f_{{\rm O}\,\rm \textsc{i}} f_{{\rm CH_{\rm x}}} n_{\rm H} \,
    - k_{\gamma, {\rm CH_{\rm x}}} \, f_{{\rm CH_{\rm x}}}
    \,=\, 0 \, .
\end{aligned}
\end{equation}
These give
\begin{equation}\label{eq:ohx2}
    f_{{\rm OH_{\rm x}}}=\frac{\frac{1}{3} k_{{\rm O}\,\rm \textsc{i}, {\rm H_3^+}} f_{{\rm O}\,\rm \textsc{i}} f_{{\rm H_3^+}}}
    {k_{{\rm C}\,\rm \textsc{i}, {\rm OH_{\rm x}}} f_{\rm{C}\,\textsc{i}}+k_{\gamma, {\rm OH_{\rm x}}} / n_{\rm H}}  \, ;
\end{equation}
\begin{equation}\label{eq:chx2}
    f_{{\rm CH_{\rm x}}}=\frac{\frac{1}{3} k_{{\rm C}\,\rm \textsc{i}, {\rm H_3^+}} f_{\rm{C}\,\textsc{i}} f_{{\rm H_3^+}} \,+\, \frac{1}{2} k_{{\rm C^+}, {\rm H_2}} f_{{\rm C^+}} f_{{\rm H_2}}}  
    {k_{{\rm O}\,\rm \textsc{i}, {\rm CH_{\rm x}}} f_{{\rm O}\,\rm \textsc{i}} +k_{\gamma, {\rm CH_{\rm x}}}/ n_{\rm H}} \,.
\end{equation}
%\FloatBarrier
\section{Cell-averaged rate equations}
\label{appendixB}
Obtaining the differential equation for the evolution of each species within a given region (e.g. a cell in a cosmological simulation) requires calculating a  PDF-weighted integral of each term on the right-hand side in Eqs. (\ref{eq:de_h2})-(\ref{eq:de_cp}). 
This leads to the following set of differential equations for the cell-averaged abundances:
\begin{multline}
    \frac{\mathrm{d} \langle f_{{\rm H_2}} \rangle }{\mathrm{d} t}\,=\,
    \langle  2k_{{\rm H_2},{\rm gr}}\,f_{{\rm H}\,\rm \textsc{i}}\,n_{\rm H}\rangle \,-\,
    \langle  k_{\gamma,{\rm H_2}}\,f_{{\rm H_2}}\rangle \,-\,
    \langle  k_{{\rm c.r.,{\rm H_2}}}\,f_{{\rm H_2}} \rangle \\ \,-\,
    \langle  n_{\rm H}\,k_{{\rm C^+},{\rm H_2}}\,f_{{\rm C^+}}\,f_{{\rm H_2}}\,\alpha_{{\rm CH_{\rm x}}}\rangle \,+\,
    \left \langle \frac{2}{3}\,n_{\rm H}\,k_{{\rm e^-},{\rm H_3^+}}\,f_{{\rm e^-}}\,f_{{\rm H_3^+}} \right \rangle \, \\
    %\,=\, 
    \underbrace{\int_{0}^{n_{\rm crit, H_2}} 2k_{{\rm H_2},{\rm gr}} \, n_{\rm H} \, \mathcal{P_{\rm M}}(n_{\rm H}) \, \mathrm{d} n_{\rm H}}_{A_1}
    \,-\, \underbrace{\int_{n_{\rm crit, H_2}}^{\infty} k_{\gamma,{\rm H_2}} \, \mathcal{P_{\rm M}}(n_{\rm H}) \, \mathrm{d} n_{\rm H}}_{A_2} \\
    \,-\, \underbrace{\int_{n_{\rm crit, H_2}}^{\infty} k_{{\rm c.r.,{\rm H_2}}} \, \mathcal{P_{\rm M}}(n_{\rm H}) \, \mathrm{d} n_{\rm H}}_{A_3} 
    \,-\, \underbrace{\int_{n_{\rm crit, H_2}}^{n_{\rm crit, C \, \textsc{i}}} n_{\rm H}\,k_{{\rm C^+},{\rm H_2}}\,f_{\rm C, tot}\,\alpha_{{\rm CH_{\rm x}}}\, \mathcal{P_{\rm M}}(n_{\rm H}) \, \mathrm{d} n_{\rm H}}_{A_4} \\
    \,+\, \underbrace{\int_{n_{\rm crit, H_2}}^{\infty} \, \frac{2}{3}\,n_{\rm H}\,k_{{\rm e^-},{\rm H_3^+}}\,f_{{\rm e^-}}\,f_{{\rm H_3^+}} \,
    \mathcal{P_{\rm M}}(n_{\rm H}) \, \mathrm{d} n_{\rm H}}_{A_5}
    \, ;
    \label{eq:de_h2_cell}
\end{multline}
%------------------------------------------------------------------
\begin{multline}
     \frac{\mathrm{d} \langle f_{{\rm CO}} \rangle}{\mathrm{d} t}\,=\,
     \left \langle \frac{1}{3}\,n_{\rm H}\,k_{{\rm O}\,\rm \textsc{i},{\rm H_3^+}}\,f_{{\rm O}\,\rm \textsc{i}}\,f_{{\rm H_3^+}}\,\alpha_{{\rm OH_{\rm x}}}  \right \rangle\,+\,
     \left \langle \frac{1}{3}\,n_{\rm H}\,k_{{\rm C}\, \textsc{i},{\rm H_3^+}}\,f_{\rm{C}\,\textsc{i}}\,f_{{\rm H_3^+}}\,\alpha_{{\rm CH_{\rm x}}}  \right \rangle \\ \,+\,
     \left \langle \frac{1}{2}\,n_{\rm H}\,k_{{\rm C^+},{\rm H_2}}\,f_{{\rm C^+}}\,f_{{\rm H_2}}\,\alpha_{{\rm CH_{\rm x}}} \right \rangle\,-\,
     \langle k_{\gamma,{\rm CO}}\,f_{{\rm CO}}  \rangle\, \\
     \,=\, \underbrace{\int_{n_{\rm crit, H_2}}^{\infty} \frac{1}{3} \, n_{\rm H}\,k_{{\rm O \, \textsc{i}},{\rm H_3^+}}\,f_{\rm O, \textsc{i}}\,f_{\rm H_3^+} \, \alpha_{{\rm CH_{\rm x}}}\, \mathcal{P_{\rm M}}(n_{\rm H}) \, \mathrm{d} n_{\rm H}}_{B_1} \\
     \,+\, \underbrace{\int_{{\rm max}(n_{\rm crit, H_2}, n_{\rm crit, C \, \textsc{i}})}^{n_{\rm crit, CO }}  \frac{1}{3} \, n_{\rm H}\,k_{{\rm C \, \textsc{i}},{\rm H_3^+}}\,f_{\rm C, tot}\,f_{\rm H_3^+} \, \alpha_{{\rm CH_{\rm x}}}\, \mathcal{P_{\rm M}}(n_{\rm H}) \, \mathrm{d} n_{\rm H}}_{B_2} \\
    \,+\, \underbrace{\int_{n_{\rm crit, H_2}}^{n_{\rm crit, C \, \textsc{i}}} \frac{1}{2} \, n_{\rm H}\,k_{{\rm C^+},{\rm H_2}}\,f_{\rm C, tot}\,\alpha_{{\rm CH_{\rm x}}}\, \mathcal{P_{\rm M}}(n_{\rm H}) \, \mathrm{d} n_{\rm H}}_{B_3} \\
    \,-\, \underbrace{\int_{n_{\rm crit, CO}}^{\infty} k_{\gamma,{\rm CO}}\,f_{{\rm C, tot}} \, \mathcal{P_{\rm M}}(n_{\rm H}) \, \mathrm{d} n_{\rm H}}_{B_4} \, ;
    \label{eq:de_co_cell}
\end{multline}
%------------------------------------------------------------------------
\begin{multline}
    \frac{\mathrm{d} \langle f_{{\rm C^+}} \rangle }{\mathrm{d} t}\,=\,
    \,-\,\langle n_{\rm H}\,k_{{\rm C^+},{\rm e^-}}\,f_{{\rm C^+}}\,f_{{\rm e^-}} \rangle
    \,-\,\langle n_{\rm H}\,k_{{\rm C^+},{\rm e^-},{\rm gr}}\,f_{{\rm C^+}}\,f_{{\rm e^-}} \rangle\\ \,-\,
    \langle \frac{1}{2} \, n_{\rm H}\,k_{{\rm C^+},{\rm H_2}}\,f_{{\rm C^+}}\,f_{{\rm H_2}}\,\alpha_{{\rm CH_{\rm x}}} \rangle\,+\,
    \langle k_{\gamma,\rm{C}\,\textsc{i}}\,f_{\rm{C}\,\textsc{i}} \rangle  \, \\
    \,=\, -\underbrace{\int_{0}^{n_{\rm crit, C \, \textsc{i}}} \, n_{\rm H} \, (k_{\rm C^+, e^-}+k_{\rm C^+, e^-, gr}) \, f_{{\rm C, tot}} \, f_{{\rm e^-}}\,\mathcal{P_{\rm M}}(n_{\rm H}) \, \mathrm{d} n_{\rm H}}_{C_1} \\
    \,-\, \underbrace{\int_{n_{\rm crit, H_2}}^{n_{\rm crit, C \, \textsc{i}}} \frac{1}{2} \, n_{\rm H}\,k_{{\rm C^+},{\rm H_2}}\,f_{\rm C, tot}\,\alpha_{{\rm CH_{\rm x}}}\, \mathcal{P_{\rm M}}(n_{\rm H}) \, \mathrm{d} n_{\rm H}}_{C_2} \\
    \,+\, \underbrace{\int_{n_{\rm crit, C \, \textsc{i}}}^{n_{\rm crit, CO }} k_{\gamma,{\rm C \, \textsc{i}}}\,f_{{\rm C, tot}} \, \mathcal{P_{\rm M}}(n_{\rm H}) \, \mathrm{d} n_{\rm H}}_{C_3} \, .
    \label{eq:de_cp_cell}
\end{multline}
%---------------------------------

In the above equations, $n_{\rm crit, H_2}$, $n_{\rm crit, C \, \textsc{i}}$, and $n_{\rm crit, CO}$ are the critical densities for the $\rm \textsc{Hi} \rightarrow H_2 $, $\rm C^+ \rightarrow C $, and $\rm C \, \textsc{i} \rightarrow CO $ transitions, respectively in a given cell (see Sect.~\ref{sec:shielding_fn} and Appendix~\ref{appendixC}). The $f_{\rm H_3^+}$ in these equations can be obtained from  Eq. (\ref{eq:h3p2}). We note that the limits of integrals $A_5$ and $B_1$ are from $n_{\rm crit, H_2}$ to $\infty$ as the formation of $\rm H_3^+$ relies on the presence of $\rm  H_2$ (see Eq.~\ref{eq:h3p1}), but the exact expression for $f_{\rm H_3^+}$ would be different for $n_{\rm H} < n_{\rm crit, C \, \textsc{i}}$  (where $f_{\rm C^+} = f_{\rm C, tot}$, $f_{\rm C \, \textsc{i}} = f_{\rm CO} = 0$), $n_{\rm crit, C \, \textsc{i}} \leq n_{\rm H} < n_{\rm crit, CO}$  (where $f_{\rm C \, \textsc{i}} = f_{\rm C, tot}$, $f_{\rm C^+} = f_{\rm CO} = 0$), and $n_{\rm H} \geq n_{\rm crit, CO}$  (where $f_{\rm CO} = f_{\rm C, tot}$,  $f_{\rm C^+} = f_{\rm C \, \textsc{i}} = 0$). Furthermore, the integrals $A_4$, $B_3$ and $C_2$ will vanish for a given cell if $n_{\rm crit, H_2} > n_{\rm crit, C \, \textsc{i}}$.

In Eqs. (\ref{eq:de_h2_cell})-(\ref{eq:de_cp_cell}), the resulting system of coupled differential equations is solved using the implicit integrator DASSL\footnote{\url{https://www.osti.gov/servlets/purl/5882821}} \citep{dassl}. DASSL uses a variable order (between 1 and 5) backward differential formula to compute the solution of the coupled ODEs after one time step. 

\subsection{Branching ratios}
\label{sec:branching_ratios}
Often there are multiple outcomes for the reaction between two species. The probability for a given outcome is represented as a branching ratio for that outcome and it denotes the fraction of times that particular outcome will occur. For example, the reaction between $\rm C^+$ and $\rm H_2$ leads to the formation of $\rm CH_2^+$, which reacts with $\rm H_2$ to give $\rm CH_3^+$. $\rm CH_3^+$ reacts with an electron to give $\rm  CH$ or $\rm  CH_2$ in $70 \%$ of the cases (these two species are referred to as $\rm  CH_{\rm x}$ in reaction 1 in Table~\ref{tab:chem_reactions}).  In the remaining $30 \%$ of the cases, $\rm CH_3^+ + e^-$ gives $\rm C \, + \, 2 H$ (reaction 2). The reaction rates in Table~\ref{tab:chem_reactions} account for these branching ratios.

\section{Calculation of \lowercase{$n_{\rm crit}$}}
\label{appendixC}
The hydrogen in a region can exist in atomic ($ {\rm H}\,\rm \textsc{i}$) and molecular (${\rm H_2}$) forms and their mean densities are related to the mean density $\langle n_{\rm H} \rangle$ of ${\rm H}$ nuclei in the cell as 
\begin{equation}\label{eq:sum_Hden}
    \langle n_{\rm H \, \textsc{i}} \rangle \,+\, 2\, \langle n_{\rm H_2} \rangle \,=\, \langle n_{\rm H} \rangle \, .
\end{equation}
The mean ${\rm H_2}$ fraction in a region is defined as
\begin{equation}
    f_{\rm H_2} = 2 \frac{\langle n_{\rm H_2} \rangle}{\langle n_{\rm H} \rangle} \, .
\end{equation}

Assuming that hydrogen shows a sharp transition from fully atomic to fully molecular at $n_{\rm H} = n_{\rm crit, H_2}$, Eq. (\ref{eq:sum_Hden}) becomes
\begin{equation}
     \int_{0}^{n_{\rm crit, H_2}} \mathcal{P_{\rm M}} \, \mathrm{d}n_{\rm H} 
    \,+\, 
    \underbrace{\int_{n_{\rm crit, H_2}}^{\infty} \mathcal{P_{\rm M}} \, \mathrm{d}n_{\rm H}}_{2 \, \langle n_{\rm H_2} \rangle / \langle n_{\rm H} \rangle} 
    \,= 1 .
    \label{eq:ncrit}
\end{equation}

For a log-normal $\mathcal{P_{\rm M}}$, this becomes
 \begin{equation}
    \frac{1}{2}\, \left[ 1+ \mathrm{Erf} \left( \frac{\ln n_{\rm crit, H_2} - \mu }{\sqrt{2}\,\sigma}\right) \right] 
    \,+\, 2 \frac{\langle n_{\rm H_2} \rangle}{\langle n_{\rm H} \rangle } = 1.
\end{equation}
If $\langle n_{\rm H} \rangle$ and $f_{{\rm H_2}}$ are known quantities, then the above equation can be written as
\begin{equation}
    f_{{\rm H_2}} = \frac{1}{2}\, \left[ 1 - \mathrm{Erf} \left( \frac{\ln n_{\rm crit, H_2} - \mu}{\sqrt{2}\,\sigma}\right) \right] \, ,
    \label{eq:ncrit_lognormal}
\end{equation}
where $\mu = \langle n_{\rm H} \rangle\,+\,0.5\sigma^2$. The critical density $n_{\rm crit, H_2}$ can be obtained by finding the root of the above equation. This equation is equivalent to equation (10) in T15. 

For a log-normal+power-law PDF, 
Eq.(\ref{eq:ncrit}) becomes
\begin{equation}
    \left[ \int_{0}^{n_{\rm trans}} \mathcal{P_{\rm M1}} \, \mathrm{d}n_{\rm H} \,+\, 
    \int_{n_{\rm trans}}^{n_{\rm crit, H_2}} \mathcal{P_{\rm M2}} \, \mathrm{d}n_{\rm H} 
    \right ] 
    \,+\, 2 \frac{\langle n_{\rm H_2} \rangle}{\langle n_{\rm H} \rangle} \,=\, 1 \, .
    \label{eq:ncrit_lognormal+powerlaw}
\end{equation}
We note that the second term in the brackets will be zero for $n_{\rm crit, H_2} < n_{\rm trans}$. While using an iterative root-finding method such as the Newton-Raphson method, one needs to evaluate the terms in the brackets differently depending on whether a given guess value for the root $n_{\rm crit, H_2}$ is smaller or larger than $n_{\rm trans}$. Thus, in this case, it is not possible to obtain an analytical expression relating $\langle n_{\rm H} \rangle$, $f_{{\rm H_2}}$, and $n_{\rm crit, H_2}$.     

%\FloatBarrier
\section{Effect of the CRIR on chemistry}
\label{appendixD}
\begin{figure}[b]
    \centering
    \includegraphics[width=0.49\textwidth,trim={0 0 0 1cm},clip]{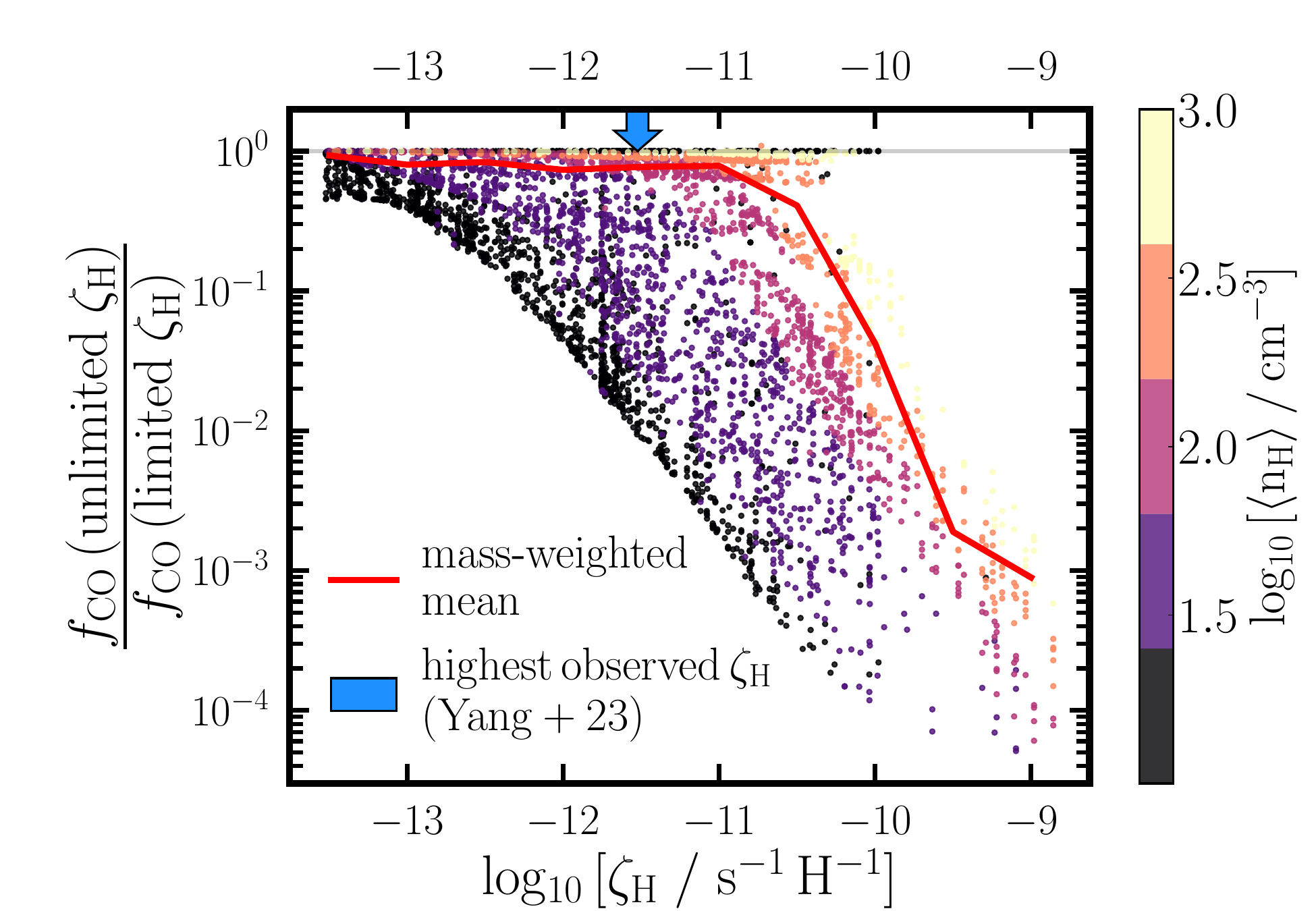}
    \caption{Ratio of the $f_{\rm CO}$ obtained when using a $\zeta_{\rm H} \propto \chi^2$without and with an upper limit on $\zeta_{\rm H}$, as a function of $\zeta_{\rm H}$ in the post-processed galaxy. Here we only show the grid cells with $\zeta_{\rm H}$ greater than the imposed upper limit of $  3 \times 10^{-14} \, \rm s^{-1} \, H^{-1}$, since the two $f_{\rm CO}$ are identical for lower CRIR values. Each grid cell is colour-coded by the mean hydrogen density $\langle n_{\rm H} \rangle$ in the cell. The red line shows the mass-weighted mean of the cells in each $\zeta_{\rm H}$ bin. We also indicate by a blue arrow the highest observed $\zeta_{\rm H}$ of $3 \times 10^{-12} \, \rm s^{-1} \, H^{-1}$ \citep[from][]{yang23}.}
    \label{fig:fco_ratio}
\end{figure}
\begin{figure*}
    \centering
    \includegraphics[width=0.96\textwidth,trim={2cm 0 6cm 6cm},clip]
    {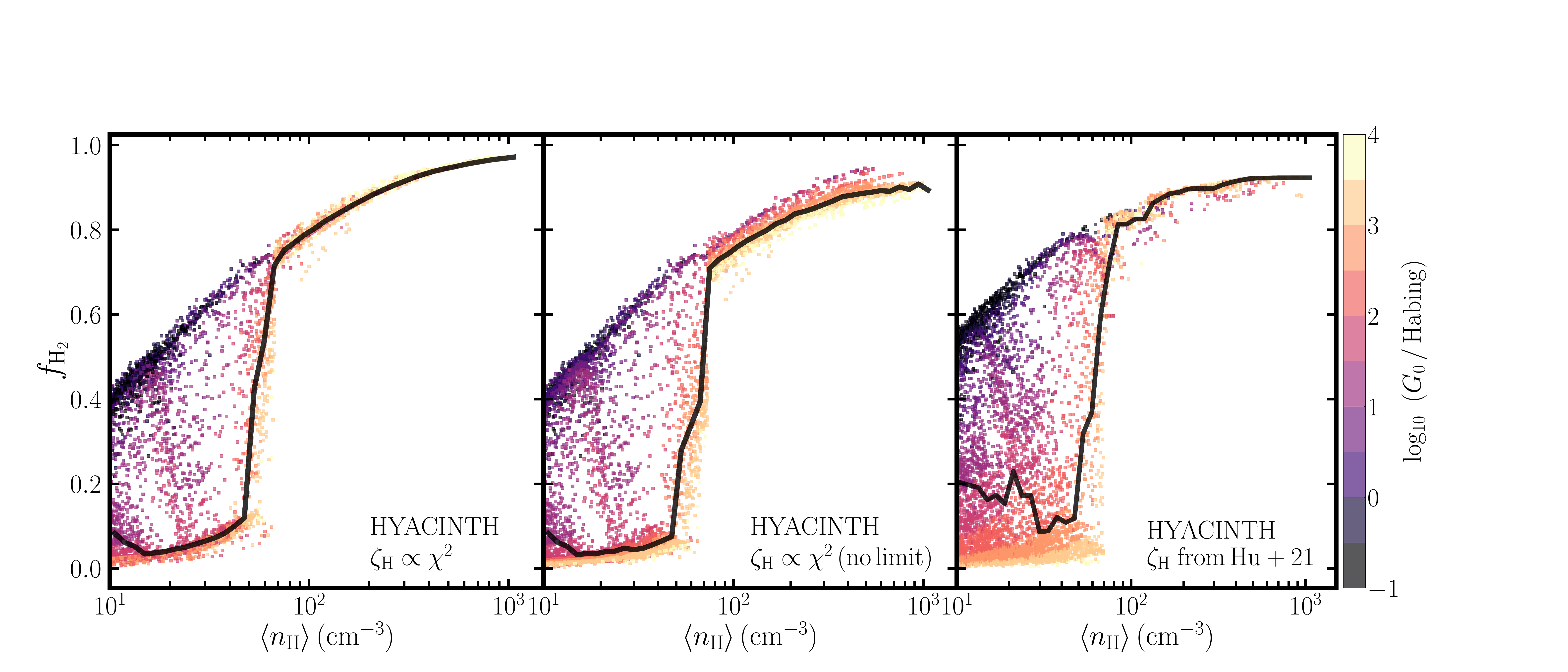}    
    \caption{Same as the left panel of Fig.~\ref{fig:h2_fraction}, but for a different CRIR used in post-processing (indicated in each panel). The  solid black line shows the median $f_{\rm H_2}$ in a given $\langle n_{\rm H} \rangle$ bin.}
    \label{fig:scatter_h2_fraction}
\end{figure*}
\begin{table*}[]
    \caption{Total mass of the different chemical species in the post-processed galaxy for different choices of the CRIR. For reference, the (dynamically evolved) $M_{\rm H_2}$ from the simulation is $4.21 \times 10^{10} \, \rm M_{\odot}$.}
    \centering
    \begin{tabular}{c c c c c c}
     \hline\hline
    \textbf{Method} & $\boldsymbol{\zeta_{\mathrm H}}$ &$\mathbf{M_{\rm H_2}}$
    & $\mathbf{M_{\rm CO}} $
    &  $\mathbf{M_{\rm C \, \textsc{i}}} $
    & $\mathbf{M_{\rm C^+}} $\\
    & &$(10^{10} \, \rm M_{\odot})$& $(10^{7} \, \rm M_{\odot})$&
    $(10^{7} \, \rm M_{\odot})$ & $(10^{7} \, \rm M_{\odot})$\\
     \hline
         HYACINTH & $\zeta_{{\mathrm{H, \,MW}}}$ & 4.70  & 9.54 & 0.63 & 0.62\\
         HYACINTH & $ \mathrm{default}$ & 4.12 & 9.38 & 0.66 & 0.66 \\
         HYACINTH & $\propto \chi^2$ (no ceiling) & 3.84 & 6.52& 0.90& 1.64  \\
         HYACINTH & \cite{hu21} & 4.08 & 9.38& 0.67 & 0.65\\
      \hline
    \end{tabular}
    \label{tab:masses}
\end{table*}

In order to investigate the impact on the chemistry of our assumptions regarding the scaling of the CRIR,
we extend the results presented in Sect.~\ref{sec:results} by repeating the post-processing of the simulated galaxy and considering different options. In Fig.~\ref{fig:fco_ratio}, we plot the ratio of the CO abundance within a cell without and with the upper limit on the CRIR introduced in  Sect.~\ref{sec:cr}.
At a given $\zeta_{\rm H}$, the effect of imposing an upper bound on $\zeta_{\rm H}$ is strongest at low densities and decreases with increasing density. Consequently, the mass-weighted ratio of the CO abundance in the two cases is very close to 1 for CRIR values up to $10^{-11} \, \rm s^{-1} \, H^{-1}$, which is larger than the highest observational estimate\footnote{Till date and to the best of our knowledge.} of $3 \times 10^{-12} \, \rm s^{-1} \, H^{-1}$ reported in \citet[][denoted by a blue arrow in Fig.~\ref{fig:fco_ratio}]{yang23}.

In Fig.~\ref{fig:scatter_h2_fraction}, we show the $f_{\rm H_2}$ obtained in each cell when using a variable $\zeta_{\rm H}$ in post-processing. We consider three different CRIRs: (i) $\zeta_{{\rm H}} \propto \chi^2$, with an upper limit of $3 \times 10^{-14} \, \rm s^{-1} \, H^{-1}$ on $\zeta_{{\rm H}}$ (left panel); (ii) $\zeta_{{\rm H}} \propto \chi^2$ without any upper limit (middle panel); (iii) the $\zeta_{{\rm H}}- \chi$ relation from \citet[][right panel]{hu21}. The median $f_{\rm H_2}$ in each case is shown by a solid black line (also shown in the left panel of Fig.~\ref{fig:h2_fraction}). We find slight variations among the three cases for the individual cells. Nevertheless, at low densities ($\langle n_{\rm H} \rangle \lesssim 50 \, \rm cm^{-3}$), $f_{\rm H_2}$ decreases with an increase in $G_0$ (the UV field in LW bands in Habing units, denoted by the colour of the points).

Finally, in Table~\ref{tab:masses}, we report the total mass of the different chemical species in the post-processed galaxy. Within the accuracy that can be expected from our simplified calculations, HYACINTH provides stable predictions for the masses that are not influenced much by the assumed scaling between the CRIR and the UV flux. The quadratic scaling with no upper limit gives the most discrepant results, with a 30\% reduction in the CO mass compensated by an increase in the masses of neutral and ionised atomic carbon. In contrast, the $\rm H_2$ mass is reduced only by $\sim 7\%$. The results obtained with our default choice and the CRIR from \cite{hu21} are nearly identical.

\end{appendix}

\end{document}